\documentclass[useAMS,usenatbib]{mn2e}
\usepackage[a4paper]{geometry}
\usepackage{graphicx}
\usepackage{color,ulem}

\def\degK{\,{}^\circ\mbox{K}}

\def\km{\mbox{\,km}}

\def\kpc{\mbox{\,kpc}}

\def\Myr{\mbox{\,Myr}}
\def\Gyr{\mbox{\,Gyr}}
\def\muG{\,\mu\mbox{G}}
\def\kms{\mbox{\,km s}^{-1}}

\def\pcc{\mbox{\,cm}^{-3}}

\def\et{et~al. }
\def\eg{e.g.,}
\def\kmskpc{{\kms\kpc^{-1}}}
\def\deg{{^\circ}}

\newcommand\aj{{AJ}}%
\newcommand\apj{{ApJ}}%
\newcommand\apjl{{ApJ}}%
\newcommand\apjs{{ApJS}}%
\newcommand\aap{{A\&A}}%
\newcommand\aaps{\ref@jnl{A\&AS}}%
\newcommand\azh{\ref@jnl{AZh}}%
\newcommand\baas{\ref@jnl{BAAS}}%
\newcommand\caa{\ref@jnl{Chinese Astron. Astrophys.}}%
\newcommand\cjaa{\ref@jnl{Chinese J. Astron. Astrophys.}}%
\newcommand\icarus{\ref@jnl{Icarus}}%
\newcommand\jcap{\ref@jnl{J. Cosmology Astropart. Phys.}}%
\newcommand\jrasc{\ref@jnl{JRASC}}%
\newcommand\memras{\ref@jnl{MmRAS}}%
\newcommand\mnras{{MNRAS}}%
\newcommand\na{\ref@jnl{New A}}%
\newcommand\nar{\ref@jnl{New A Rev.}}%
\newcommand\pra{\ref@jnl{Phys.~Rev.~A}}%
\newcommand\prb{\ref@jnl{Phys.~Rev.~B}}%
\newcommand\prc{\ref@jnl{Phys.~Rev.~C}}%
\newcommand\prd{\ref@jnl{Phys.~Rev.~D}}%
\newcommand\pre{\ref@jnl{Phys.~Rev.~E}}%
\newcommand\prl{\ref@jnl{Phys.~Rev.~Lett.}}%
\newcommand\pasa{\ref@jnl{PASA}}%
\newcommand\pasp{\ref@jnl{PASP}}%
\newcommand\pasj{\ref@jnl{PASJ}}%
\newcommand\qjras{\ref@jnl{QJRAS}}%
\newcommand\rmxaa{{Rev. Mexicana Astron. Astrofis.}}%
\newcommand\skytel{\ref@jnl{S\&T}}%
\newcommand\solphys{\ref@jnl{Sol.~Phys.}}%
\newcommand\sovast{\ref@jnl{Soviet~Ast.}}%
\newcommand\ssr{\ref@jnl{Space~Sci.~Rev.}}%
\newcommand\zap{\ref@jnl{ZAp}}%
\newcommand\nat{\ref@jnl{Nature}}%
\newcommand\iaucirc{\ref@jnl{IAU~Circ.}}%
\newcommand\aplett{\ref@jnl{Astrophys.~Lett.}}%
\newcommand\apspr{\ref@jnl{Astrophys.~Space~Phys.~Res.}}%
\newcommand\bain{\ref@jnl{Bull.~Astron.~Inst.~Netherlands}}%
\newcommand\fcp{\ref@jnl{Fund.~Cosmic~Phys.}}%
\newcommand\gca{\ref@jnl{Geochim.~Cosmochim.~Acta}}%
\newcommand\grl{\ref@jnl{Geophys.~Res.~Lett.}}%
\newcommand\jcp{\ref@jnl{J.~Chem.~Phys.}}%
\newcommand\jgr{\ref@jnl{J.~Geophys.~Res.}}%
\newcommand\jqsrt{\ref@jnl{J.~Quant.~Spec.~Radiat.~Transf.}}%
\newcommand\memsai{\ref@jnl{Mem.~Soc.~Astron.~Italiana}}%
\newcommand\nphysa{\ref@jnl{Nucl.~Phys.~A}}%
\newcommand\physrep{\ref@jnl{Phys.~Rep.}}%
\newcommand\physscr{\ref@jnl{Phys.~Scr}}%
\newcommand\planss{\ref@jnl{Planet.~Space~Sci.}}%
\newcommand\procspie{\ref@jnl{Proc.~SPIE}}%

\title[The Galactic Branches as Evidence for Transient Spiral Arms]
      {The Galactic Branches as a Possible Evidence for Transient Spiral
        Arms}

\author[P\'erez-Villegas, G\'omez \& Pichardo]
{Angeles P\'erez-Villegas$^{1,2}$ \thanks{E-mail: a.perez@crya.unam.mx},
Gilberto C. G\'omez$^{1}$ \thanks{E-mail: g.gomez@crya.unam.mx} 
and B\'arbara Pichardo$^3$ \thanks{E-mail: barbara@astro.unam.mx}
\\
$^{1}$Centro de Radioastronom\'ia y Astrof\'isica,
      Universidad Nacional Aut\'onoma de M\'exico, Apdo. postal 3-72, Morelia Mich. 58089, M\'exico \\
$^{2}$ Max-Planck-Instit\"ut f\"ur Extraterrestrische Physik, Gie\ss enbachstra\ss e, D-85741 Garching, Germany\\    
$^{3}$Instituto de Astronom\'\i a, Universidad Nacional Aut\'onoma de M\'exico, Apdo. postal 70-264 Ciudad Universitaria, D.F. 04510, M\'exico}

\begin{document}

\date{Accepted . Received ; in original form }

\pagerange{\pageref{firstpage}--\pageref{lastpage}} \pubyear{}

\maketitle

\label{firstpage}

\begin{abstract}
With the use of a background Milky-Way-like potential model, we
performed stellar orbital and magnetohydrodynamic (MHD)
simulations. As a first experiment, we studied the gaseous response to
a bisymmetric spiral arm potential: the widely employed cosine
potential model and a self-gravitating tridimensional density
distribution based model called PERLAS. Important differences are
noticeable in these simulations, while the simplified cosine potential
produces two spiral arms for all cases, the more realistic density
based model produces a response of four spiral arms on the gaseous
disk, except for weak arms -i.e. close to the linear regime- where a
two-armed structure is formed. In order to compare the stellar and gas
response to the spiral arms, we have also included a detailed periodic
orbit study and explored different structural parameters within
observational uncertainties.  The four armed response has been
explained as the result of ultra harmonic resonances, or as shocks
with the massive bisymmetric spiral structure, among other. From the
results of this work, and comparing the stellar and gaseous responses,
we tracked down an alternative explanation to the formation of
branches, based only on the orbital response to a self-gravitating
spiral arms model. The presence of features such as branches, might be
an indication of transiency of the arms.

\end{abstract}


\section{Introduction}\label{sec:Intro}

Spiral arms are one of the the most striking, beautiful and
scientifically challenging structures of disc galaxies. They have
fascinated and intrigued astronomers for centuries. As a rough initial
approximation to treat this complex problem, the spiral arms had to be
assumed as almost massless and/or with extremely modest pitch angles,
in order to obtain a solution as a linear perturbation to the
axisymmetric background potential. The proposed solution to the
permanence of spiral arms in galaxies was based on the spiral density
wave linear theory (\citealt{LS64}, inspired in the 60's work of
B. Lindblad and P.O. Lindblad). The analytical solution at first order
of the theory, known as the tight-winding approximation (TWA), that
represents a weak potential model, extremely idealized as a smooth,
even negligible perturbation to the background potential, was modeled
as a simplified cosine function to represent the spiral arms
gravitational potential. Probably followed by this initial attempt of
solving the nature of spiral arms, a lot of work has tackled
repeatedly the problem of modeling them, as a simplified periodic
function that disregards the importance of their dynamical
effects. However, spiral arms have proven to be an influential feature
on galactic modern models (such as PERLAS and N-body models, based on
three dimensional density structures), far beyond of a simple smooth
perturbation.

If we compare the spiral arms with the galactic bar, the latter has a
mass between 10 and 20\% of the disc mass
\citep{Matsumoto_et82,Dwek95,Zhao96,WeSell99}, unlike the spiral arms
that have a mass smaller in general than 5\% of the disc mass
\citep{PMME03}. Consequently, the largest influence on the disc of
strongly barred galaxies is, of course, due to the central
bar. Therefore one would be tempted to oversimplify the scenario, by
assuming that spiral arms are not influential at all on the dynamics
of the disc. However, there are determinations that suggest that the
majority of spiral galaxies, or at least those with clearly delineated
spiral arms, are rather far from being linear
\citep{Antoja11,Antoja10,Antoja09,VSK06,Roca_et13, Roca_et14, PMME03,
  Kawata_et14, SellCal14, Sell11}.

Based on all those rigorous studies, spiral arms seem to merit an
extra effort to model them beyond a simple perturbing term. This fact
is even more significant in the case of the interstellar gas, which is
considerably more sensitive to the details of the potential, and in
general, responds strongly even if the mass of the spiral arms is
only a small fraction of the axisymmetric background
\citep{KKK14,Gomez_et13, ShuMR73}.

{\bf For almost a century, formal studies of spiral arms have been
  carried on. Yet, the spiral arms morphology, origin and nature are
  still
  poorly understood. With respect to their morphology, there is plentiful
  theoretical and observational literature on structures related to
  spiral arms, such as spurs, branches, feathers and beads,
  and their plausible explanations
  \citep{W70,L70,ShuMR73,S76,ED80,SP01,LeeShu12,Kim_Ost06,
    Chakrabarti_03,Kim_Ost02,Dobbs_et11,SO06,CS08,B88}. Regarding
  spurs and feathering, interesting scenarios to explain these
  features include plain hydrodynamic simulations, where the
  Kelvin-Helmholtz instability creates spurs produced by shocks with
  the spiral arms (\citealt{WK04}; an extension to self-gravitating
  three dimensional models of this work is presented in
  \citealt{Kim_Ost06}). Other scenarios include the use of global MHD
  simulations, adding gas self-gravity and magnetic fields that
  produce differential compression of gas flowing through the arms
  resulting in the formation of sheared structures in the interarm
  regions that resemble spurs and feathers \citep{SO06,Kim_Ost02}, or
  through gravitational or magneto-Jeans instabilities \citep{B88}.

On the other hand, features such as branches are significantly
longer than
spurs that emerge almost parallel to the main spiral arms but with
different pitch angles (usually smaller), and are generally
associated with resonances. It is worth to note that, in the
literature, the names spurs, feathers and branches are assigned to
slightly different structures.
In this work we will adopt the definition given by
\citet{Feitz_Sch82} and \citet{Chakrabarti_03}, where the ``spurs''
and ``feathers'' are small structures that arise almost perpendicularly
from the main spiral arm, and ``branches'' are narrow structures that
arise almost parallel to the main spiral arm, with smaller pitch
angles and extension longer than spurs (examples of branches
can be seen in galaxies like NGC 309, NGC 1637, NGC 2997, NGC 6946,
among others). Regarding such structures, theory and gas simulations
have already shown a bifurcation of spiral arms produced by
ultraharmonic resonances induced by the main spiral arm on the
background gas flow \citep{ShuMR73,PH94,PH97,Chakrabarti_03,AL92}. In
this case, the bifurcation of the arms is expected due to the topology
of the stable periodic orbits at the 4/1 resonance. Other
interpretations also include the possibility that branches are the
response of the gas produced by strong shocks against the main massive
spiral arms, i.e. the ones made mainly of older and smaller stars {\bf
  \citep{Fujimoto68,Roberts69,ShuMR73,BW67,RHV79, Martos_et04,
    KimKim14, Chakrabarti_03,Yanez_et_2008}.}}

Specifically, regarding the Milky Way Galaxy, its general structure
has been extensively studied and the latest determinations seem to
agree with two grand design symmetric spiral arms, seen on IR and
optical, and some other weaker arms, detectable mainly in optical
observations \citep{Drimmel_Spergel01,Benjamin_etal05,Drimmel00, 
Drimmel_Spergel01,Vallee13, Vallee02}. The weaker arms ({\bf branches}) 
have been explained through hydrodynamic simulations as the response 
to the two-armed stellar pattern. For example, \citet{EngGer99} and
 \citet{Fux99} found that the gas response to a barred potential can produce 
 a four-armed spiral structure. Likewise, from hydrodynamic and 
 magnetohydrodynamic (MHD) simulations, the gas responds to an 
 imposed two-armed spiral potential with four spiral arms 
 \citep{Gomez_et13, Martos_et04, ShuMR73}. In this context, 
 the gas component shocks at the position of the spiral
arms, producing a density enhancement and star formation
\citep{Fujimoto68, Roberts69, Moore_et12,Seo_Kim14} that, depending
on the relative velocity between the gas and the spiral arm, dwell
upstream or downstream or in the spiral arm.  In particular,
\citet{Roberts69} showed that the nonlinear response of the gas to a
stationary stellar spiral arm potential may produce two shocks, which
then would be associated to a doubling of the spiral arm number as
seen in the gas component. \citet{ShuMR73} demonstrated that shocks in
galaxies arise necessarily if the spiral arm strength exceeds a
certain critical value. Additionally, they found a range of values for
the wave frequency that generates an ultraharmonic resonance which can
provoke a secondary compression of the interstellar gas. This effect
has been related to the origin of the Carina arm in the Milky Way
\citep{ShuMR73}, for example. However, since branches have been
traditionally explained as shocks induced by the spiral arms, it is
important to mention other observational work that shows there is
little difference (or non at all) in the star formation efficiency
(i.e. no shocks) between the spiral arms regions and the rest of the
disk \citep{FRWL10, Eden_et13,Foyle_et11,Dobbs_et11}.

Finally, regarding their nature, a problem of great interest is whether
spiral arms are a transient or a long-lived feature. On this matter,
recent numerical simulations show that spiral arms are transient and
recurrent structures \citep{Dobbs_Bonnell06, Wada11,Roskar12,
  Donghia13, Perez-Villegas_et13, Perez-Villegas_et12,
  Sell11,SellCal14,Kawata_et14,Foyle_et11}. However, in a study by
\citet{Scarano_Lepine13} on a sample of 27 galaxies, the authors
concluded that the break found in the radial metallicity distribution
near to corotation resonance (CR), implies that the spiral structure
is a rather long lived feature. Other authors, employing different
techniques, seem to find observational proofs of long-lasting spiral
arms \citep{Martinez-Garcia13, Donn_Thom94,Zhang98}.

In this work we explore the gas response to a bisymmetric spiral arm
potential. For this purpose we employed two different models, a cosine
potential (generally used in literature) and the three-dimensional,
density distribution based potential PERLAS, applied to the particular
case of a Milky Way-like galaxy. With this study we find an
alternative explanation to the formation of branches in disk galaxies,
and their relation to transient or long-lived spiral arms.

This paper is organized as follows. In Section \ref{sec:method}, the
galactic potential used to compute the stellar orbits and the MHD
initial simulation set-up are described. In Section \ref{results}, we
present first a comparison between the models of the spiral arms:
PERLAS and the cosine potential; second, we present the gas response
to spiral arms PERLAS model changing the structural parameters such as
the pitch angle and the mass of the spiral arms, and their connection
with the presence of galactic branches as a signature of transient
spiral arms. Finally, we present a discussion and our conclusions in
Section \ref{sec:conclusions}.

\section{Methodology and Numerical Implementation}\label{sec:method}

Motivated by the fact that the gas in galaxies is dynamically colder
than the stellar disk, in addition to being collisional, we can expect
it to be much more sensitive to details of the given potential than
stars. Therefore, any differences between the cosine and PERLAS
potentials might be magnified in gas with respect to the stellar
response. {\bf And so,} for purposes of comparison, we produce the whole study
employing both potentials. In this section we introduce briefly the
cosine and PERLAS potential models and the MHD setup applied to a
Milky Way like Galaxy.

In all cases, the spiral arm models are superimposed to the
axisymmetric background potential of \citet{AS91}, which includes a
Miyamoto-Nagai bulge and disk, and a supermassive spherical
halo. Table \ref{tab:param} presents the basic parameters of the
axisymmetric background potential.

\begin{table*}
  \centering 
  \begin{minipage}{0.6\textwidth} 
    \caption{Parameters of the Galactic Model } 
    \begin{tabular}{@{}lcc@{}} 
      \hline 
      Parameter                                 & Value       & Reference\\ 
       \hline
 &{\it Axisymmetric Components}  \\
 \hline
      $R_0$& 8.5 kpc & 1 \\  
       $\Theta_0$ &220 km s$^{-1}$ & 1 \\ 
 Bulge mass& $1.41\times10^{10}$ M$_\odot$ &1 \\
 Disk mass &$8.56\times10^{10}$ M$_\odot$&1\\
 Halo mass&$8.002\times10^{11}$ M$_\odot$&1\\ 
      \hline 
      &{\it Spiral Arms}  \\
      \hline
      Locus               & Logarithmic & 2, 3\\
      Arms number & 2 &  4\\ 
      Pitch angle$(i)$      & $15.5\pm3.5\deg$ & 4 \\ 
      $M_{arms}/M_{disk}$&$0.03\pm0.02$& 3\\
      Mass &$2.7-5.4 \times10^{9}$ M$_\odot$& 3 \\
      Inner limit                & $3.3\kpc$  & ILR position based\\ 
      Outer limit & 12 \kpc& CR position based\\
      Scale-length       & $2.5\kpc$   & Disk based \\ 
      Pattern speed ($\Omega_{P}$)  & $-20 \kmskpc$& 5 \\ 
      \hline 
    \end{tabular} 
    \label{tab:param} 
     \\References ---(1) \citealt{AS91}; (2) \citealt{GP98}; (3) \citealt{PMME03};
     (4)\citealt{Drimmel00}; (5) \citealt{Martos_et04}.
  \end{minipage} 
 
\end{table*} 

\subsection{Spiral arm models}
\label{sec:sp_models}

\subsubsection{Cosine Potential}\label{cosine}

As mentioned before, a large majority of the investigations on the gas
associated to spiral arms model them as a linear perturbation of the
axisymmetric background, represented by

\begin{equation}
 \label{eq:cosine}
\Phi_{sp}(R,\phi)=f(R)\,\cos\left[2\phi+g(R)\right],
\end{equation}

\noindent
where $R,\phi$ are cylindrical coordinates, $f(R)$ is the amplitude
function of the perturbation, given by \citet{CG86} as
$f(R)=-ARe^{\epsilon_s R}$, where $A$ is the amplitude and
$\epsilon_s$ is the inverse of the scale-length. Finally, $g(R)$
describes the geometry of the spiral pattern ({\it locus}), given by
\citet{RHV79}, as

\begin{equation}
 \label{eq:locus}
g(R)=-\frac{2}{N\tan i_p}\ln[1+(R/R_s)^N],
\end{equation} 

\noindent where $i_p$ is the pitch angle, $R_s$ is the start position
for the spiral arms, and $N$ is a constant that shapes the starting
point of the spiral arms, {\bf so that} $N\rightarrow 0$
represents a $180\deg$ tip transition from the bar to the spiral arms,
while $N\rightarrow\infty$ means a $90\deg$ tip transition from the
bar to the spiral arms. In the present work, we set it equal to 100
(for further details, see \citealt{PMME03}).

\subsubsection{PERLAS model} \label{perlas}
In contrast, the PERLAS spiral arms model \citep{PMME03} is a
bisymmetric, three-dimensional, self-gravitating stationary model
based on an adjustable mass distribution, rather than a local arm
approximation, as the cosine potential. Several studies have shown
that there are differences in the stellar orbital dynamics when the
spiral arms are modeled with the cosine potential or with the PERLAS
model \citep{PMME03,Antoja09,Antoja11}. Furthermore, chaotic orbital
studies have demonstrated that a more detailed spiral arm potential
(\eg the PERLAS model) induces an important fraction of chaos, enough
to destroy the spiral structure \citep{Perez-Villegas_et13,
  Perez-Villegas_et12}, when all chaotic behavior was originally
attributed to effects produced by the bar, such as overlapping of
resonances \citep{Contopoulos_H12, Contopoulos67}.

PERLAS is constructed by a superposition of individual oblate
inhomogeneous spheroids along a given locus (the same than in the
cosine potential, equation \ref{eq:locus}) and superimposed to the
axisymmetric background. In this model, the spiral arms have a well
defined mass, unlike the cosine potential, where the spiral arms are
treated as a periodic function of the potential not straightforward
translatable to mass.  The mass assigned to build the spiral arms in
the PERLAS model is subtracted from the disk mass.  Thus, the
inclusion of PERLAS to the Galactic model does not modify the total
mass of the original axisymmetric background.  In Table
\ref{tab:param} we present the parameters of spiral arms that we used
in our simulations.  For further details about the PERLAS model, we
refer the reader to \citet{PMME03}.

\subsection{Force fitting} \label{sec:fit}
In order to compare PERLAS with the cosine potential, we need to fit
the spiral arm strength to make them as similar as possible.  We
achieve this by adjusting the amplitude of the cosine potential (the
factor $A$ in eq. \ref{eq:cosine}) so that the resulting radial and
azimuthal forces approximate the ones obtained using the PERLAS model.

\begin{figure*}
\includegraphics[width=\textwidth]{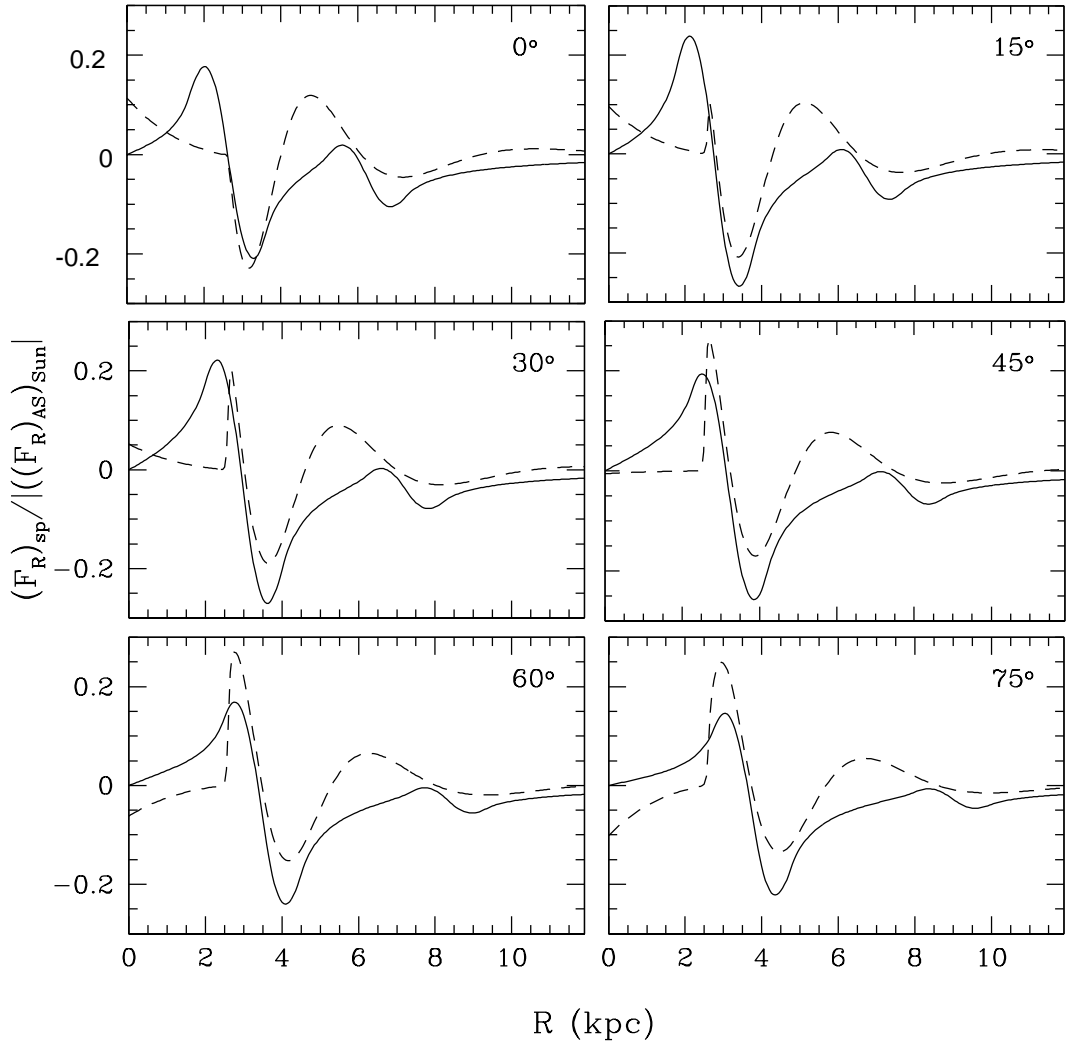}
\caption{Radial force of the spiral arms along of the different radial
  lines, starting in $0\deg$ up to $165\deg$, each $15\deg$.  The
  radial force is scaled to the axisymmetric background radial force
  at the solar position. Continuous lines represent PERLAS, dotted
  lines represent the best fit with equation \ref{eq:cosine}.}
\label{fig:frad}
\end{figure*}

\begin{figure*}
\includegraphics[width=\textwidth]{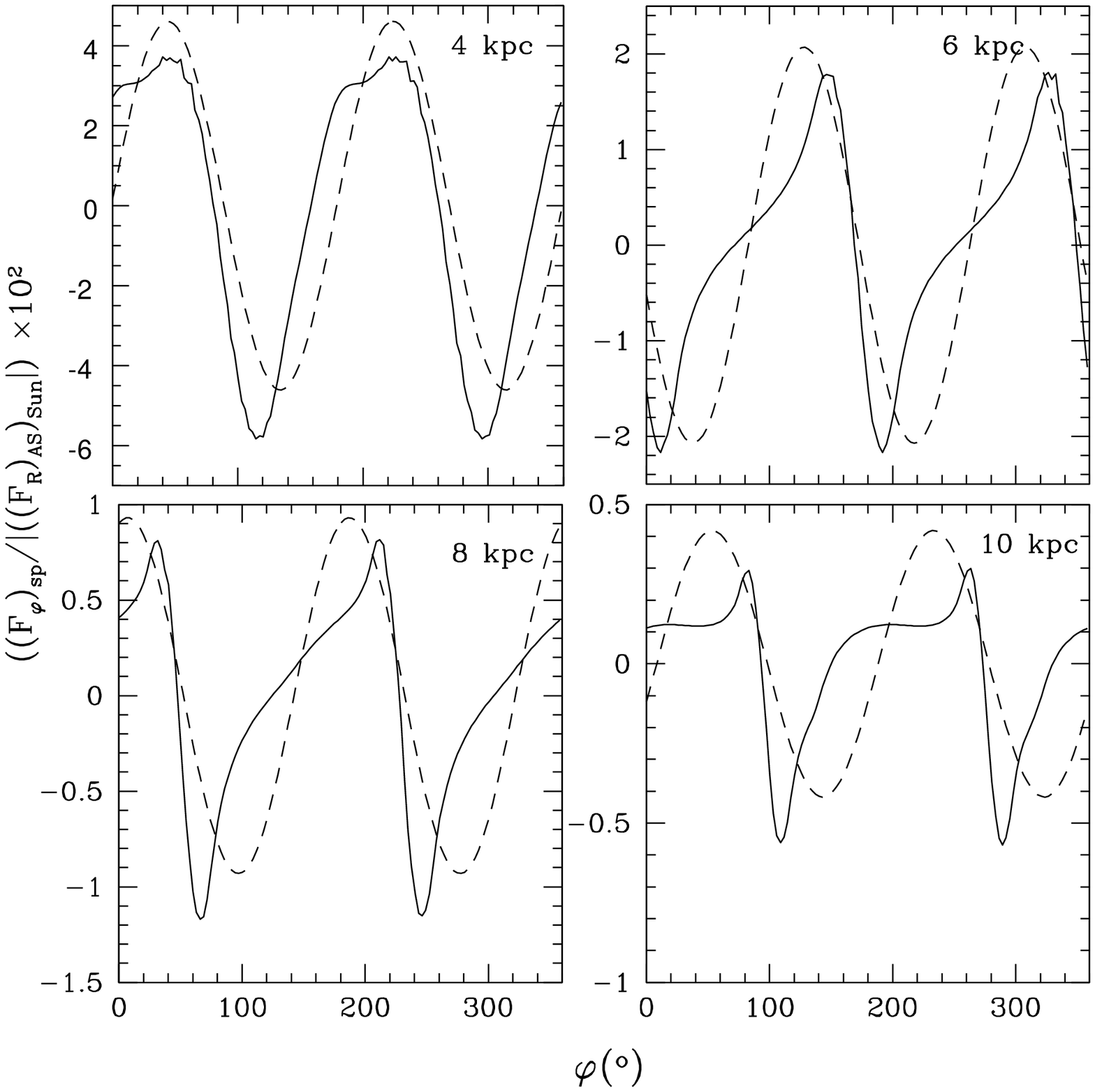}
\caption{Azimuthal force at the different radius, from $4$ to
  $10\kpc$, each $2\kpc$.  The radial force is scaled to the
  axisymmetric background radial force at the solar
  position. Continuous lines represent PERLAS, dotted lines represent
  our fit with equation \ref{eq:cosine}. }
\label{fig:fazi}
\end{figure*}

Figures \ref{fig:frad} and \ref{fig:fazi} show the resulting fit for
the case of a Milky Way-like galaxy, where the pitch angle is
$15.5\deg$ and spiral arm mass (in PERLAS) is $3\%$ of the disk mass.
In this case, the cosine amplitude $A$ is
$650\km^2\sec^{-2}\kpc^{-1}$. In Figure \ref{fig:frad}, each panel
represents different radial lines, starting in $0\deg$ up to $75\deg$.
In the Figure \ref{fig:fazi}, each panel shows the computed azimuthal
force at different radii, from $4$ to $10\kpc$.  Continuous lines
represent PERLAS, dotted lines represent our fit to the cosine
potential.

\subsection{Spiral arm strength} \label{sec:strength}

\begin{figure}
\includegraphics[width=0.5\textwidth]{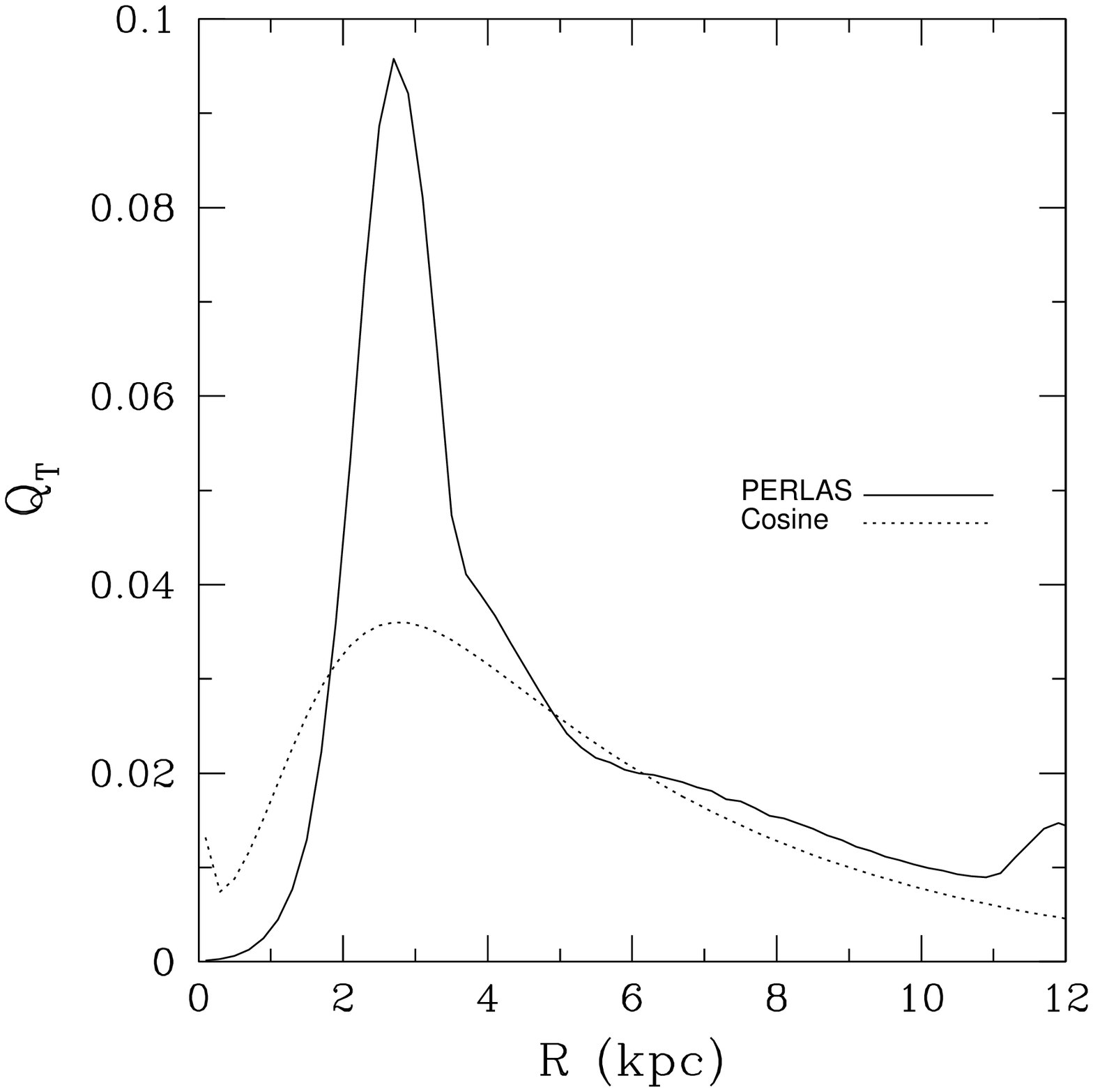}
\caption{$Q_T(R)$ parameter for the spiral arms of a Milky Way-like
  galaxy for the PERLAS ({\it solid line}) and cosine ({\it dotted
  line}) potential models.}
\label{fig:QT}
\end{figure}

The spiral strength is related to the pitch angle and the mass of the
arm. In the PERLAS model, the spiral arm mass is a small fraction of
the disk mass. To quantify the strength of the spiral arms, we
calculated the $Q_T(R)$ parameter \citep{SaTu80, ComSan81}, frequently
used to quantify the strength of bars and spiral arms
\citep{BuBl01,LS02}.  The value of $Q_T(R)$ is given by

\begin{equation}
  Q_T(R)=\frac{F_T^{\rm max}(R)}
                   {|\langle F_R(R) \rangle|},
  \label{eq:Q_T}
\end{equation}

\noindent
where $F^{\rm max}_T
=|\left(\partial\Phi(R,\theta)/\partial\theta\right)/R|_{\rm max}$
represents the maximum amplitude of the tangential force at radius
$R$, and $\langle F_R(R)\rangle$, is the average axisymmetric radial
force.  Figure \ref{fig:QT} shows $Q_T(R)$ for the PERLAS (solid line)
and cosine (dotted line) potential models, for the case where the
spiral arms mass of the PERLAS model has a 3\% of the disk mass.

\begin{figure}
\includegraphics[width=0.5\textwidth]{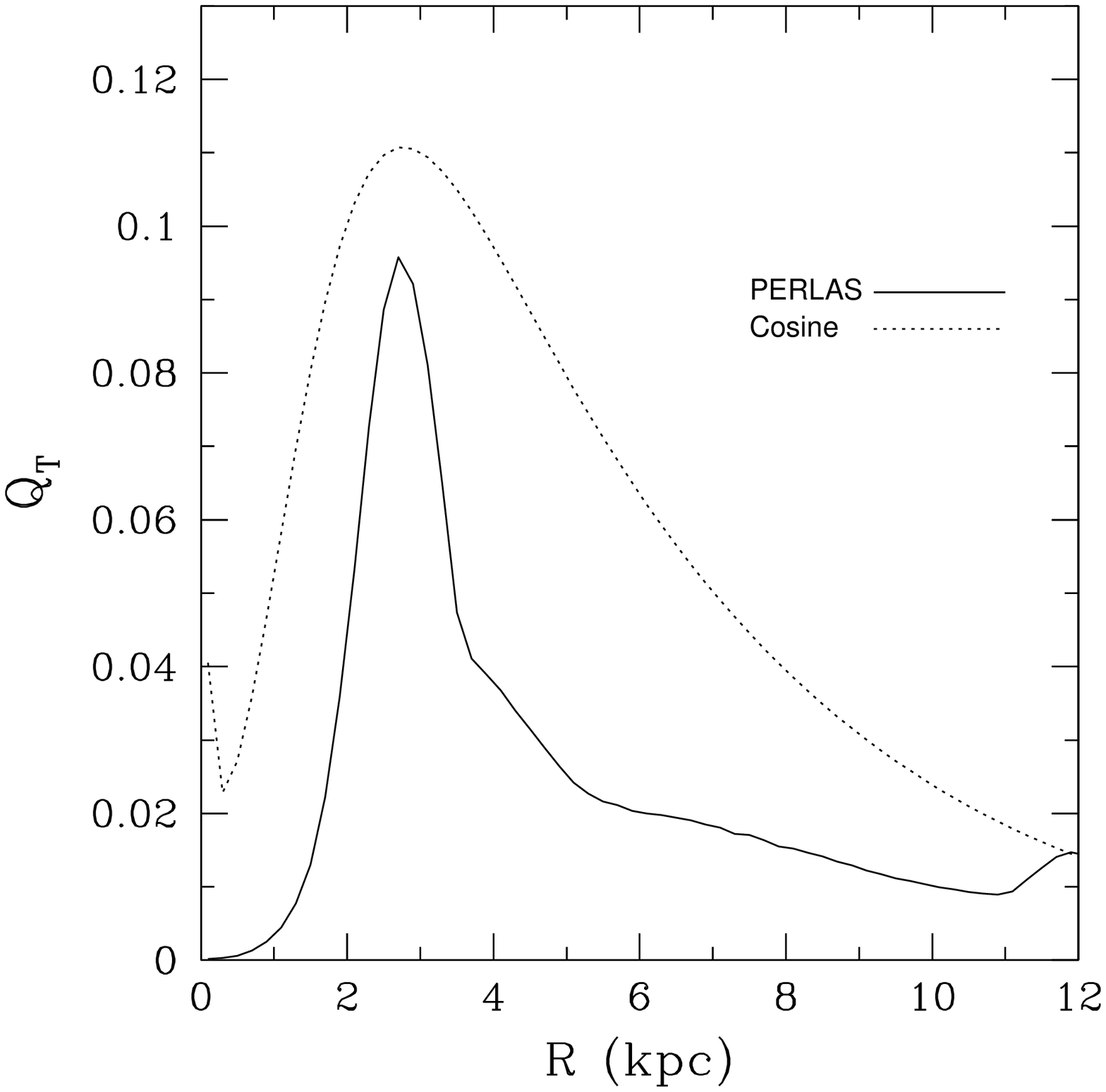}
\caption{Similar to fig. \ref{fig:QT}, with the amplitude of the cosine model
  increased.
}
\label{fig:QT_am20}
\end{figure}

As opposed to Figures \ref{fig:frad} and \ref{fig:fazi}, where the
force amplitude is almost the same, Figure \ref{fig:QT} shows that
there is a difference between both potentials.  While the maximum
value of $Q_T$ for PERLAS is $\sim 0.096$, for cosine is only $\sim
0.035$.  Therefore, if the arm strength is measured using equation
(\ref{eq:Q_T}) instead of the arm force, it is necessary to increase
the cosine model amplitude.  Figure \ref{fig:QT_am20} shows the $Q_T$
parameter corresponding to {\bf an} increased cosine amplitude, in this
case, $A=2000\km^2\sec^{-2}\kpc^{-1}$. In this case, since $Q_T$ for
the cosine model is larger than PERLAS along all the radial range,
this case should be considered as an example to test the gaseous disk
response to an extreme cosine model, in order to verify if at this
forcing the cosine can reproduce what PERLAS does. In this way we are
bracketing the cosine potential within
the values of the force for the PERLAS model.

\subsection{The MHD set-up}

The initial set-up of the simulations consisted on a
gaseous disk, with density profile given by $n(r) = n_0 \rm{exp}
[-(r - r_0)/r_d]$, where $n_0 = 1.1\pcc$, $r_0 = 8\kpc$ and
$r_d = 15\kpc$. The gas follows an isothermal equation of state with 
temperature $T = 8000\degK$.
Additionally, the gas is permeated by a magnetic field, initially in the
azimuthal direction, with an intensity given by
$B(r) = B_0 \exp [-(r - r_0)/r_B]$, where $B_0 = 5 \muG$ and
$r_B = 25\kpc$.
The disk is set up in rotational equilibrium
between the centrifugal force, the thermal and magnetic pressures,
magnetic tension and the background axisymmetric potential
\citep{AS91}.
This equilibrium is perturbed by the spiral arm potentials
under study, both rotating with a pattern speed $\Omega_p= 20\kmskpc$.

We employed the {\sc zeus} code \citep{SN92a,SN92b} to solve the MHD
equations, which is a finite difference, time explicit, operator
split, Eulerian code for ideal MHD.
We used a 2D grid in cylindrical
geometry, with $R \in [1.5, 22]\kpc$ and a full circle in the
azimuthal coordinate, $\phi$, using $750 \times 1500$ grid points.
Both boundary conditions in the radial direction were outflowing.
All calculations are performed in the reference frame of the spiral
arms. No self-gravity of the gas was considered.


\section{Results} \label{results}
We present in this section
a comparison between the
gas response to the simplified cosine potential for the spiral arms,
and the density distribution based model PERLAS. In the limit for weak
and/or small pitch angles (approximately linear regime), both models
behave very similar as expected, however, for stronger arms (more
massive or larger pitch angles), from this comparison we find severe
differences in the gas behavior and formation of spiral arms and
branch-like structures. The deviation of the response between both
models is induced by the basic differences of these potentials.
We then
present an interpretation on the presence of galactic
branches prior to the ultraharmonic 4:1 resonance and connect them to
a possible signature of transient spiral arms.

\subsection{Gas Response Comparison: PERLAS {\it vs.} cosine Potential Models}
\label{sec:comparison}
In order to compare both potentials, we fit the cosine potential with
PERLAS to make it as similar as possible in the force amplitude
(\S\ref{sec:fit}). In all cases, the axisymmetric gaseous disk,
initially in rotational equilibrium, was perturbed by the spiral
potential either PERLAS or the cosine.

\begin{figure*}
\includegraphics[width=0.4\textwidth]{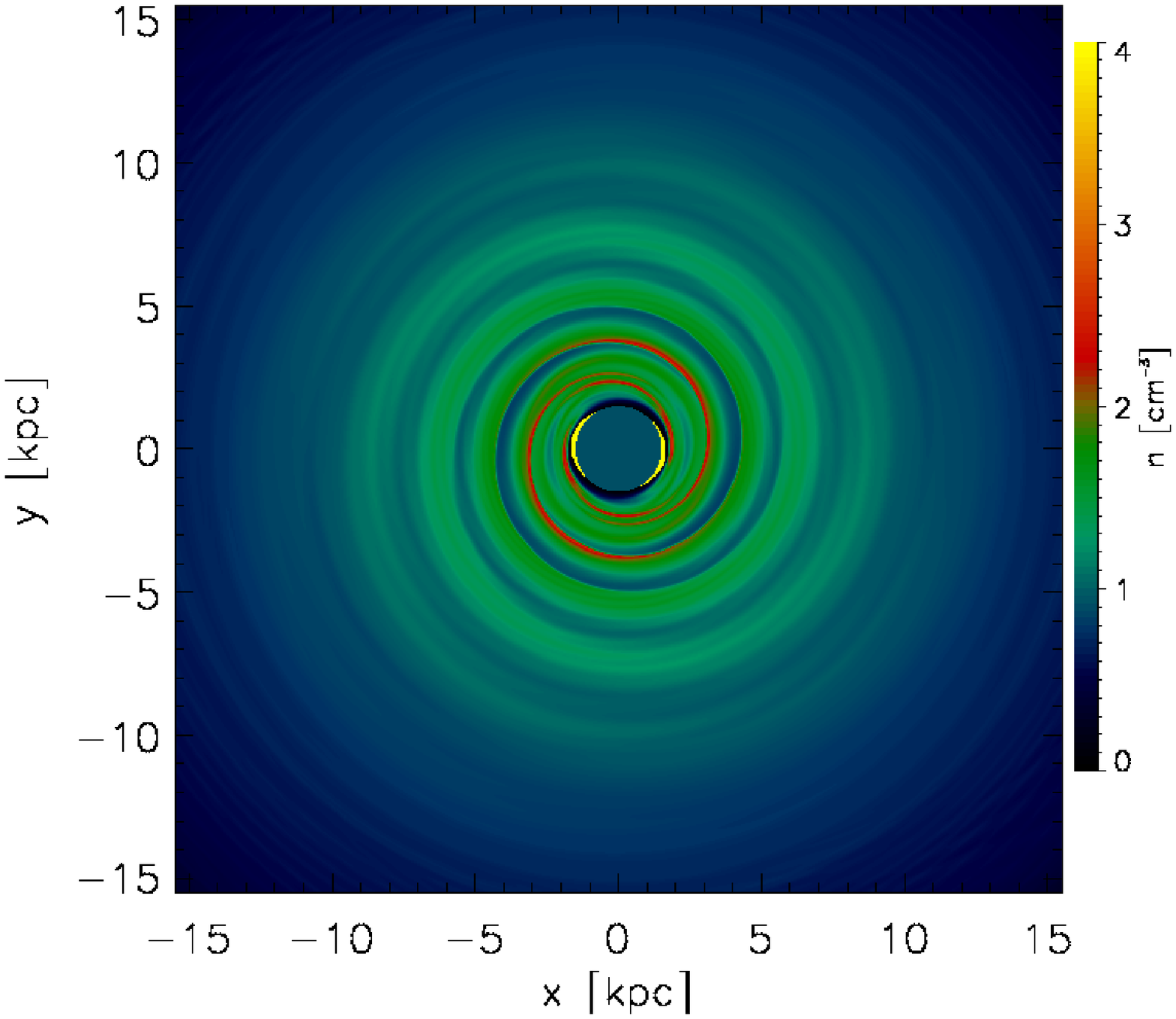}
\includegraphics[width=0.4\textwidth]{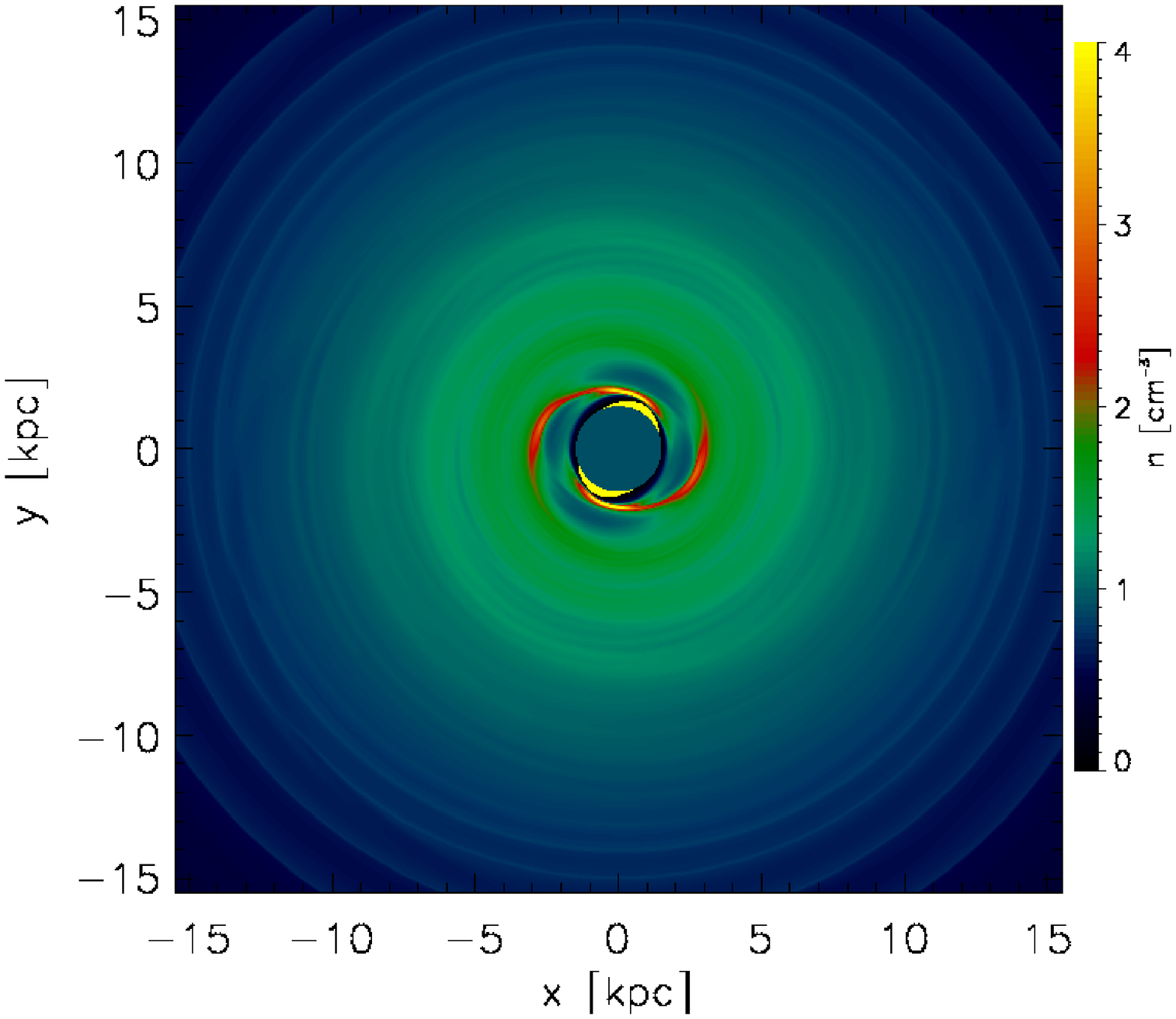}
\caption{Density distribution for the simulation with the cosine
  potential ({\it left panel}) and with the PERLAS model in the linear
  regime ({\it right panel}) after $1.2 \Gyr$.  In both cases, the
  pitch angle is $6\deg$.  The amplitude for the cosine potential is
  $A=100\km^2\sec^{-2}\kpc^{-1}$ and the spiral arm mass for the
  PERLAS model is 1\% of the stellar disk mass.}
\label{fig:gas_cos_perlas}
\end{figure*}

As mentioned already, the cosine potential represents a simple
solution from the density wave linear theory, self-consistent for
tightly wound spiral arms (TWA, i.e. where the perturbation is very
small, which means small pitch angles or with very reduced masses).
The Milky Way Galaxy and the most of spiral galaxies are actually far
from this regime. Thus, it should not be surprising that the gaseous
disk \citep{Gomez_et13} and the stellar orbits
\citep{Perez-Villegas_et12} show a different structure when subjected
to a self-gravitating, more realistic model instead of a local
approximation. With this in mind, the gaseous disk response to both
models should be similar if we focus on a region of the parameter
space where both potentials are valid, i.e., if we set PERLAS and the
cosine potential so that it is approximately in the linear regime,
with a very small pitch angle and spiral arm mass. In Figure
\ref{fig:gas_cos_perlas} we show the linear regime for both
potentials. The density distribution with the cosine potential is
presented in the left panel and PERLAS model in the right panel, the
pitch angle is $6\deg$, the spiral arms mass is 1\% of the disk mass,
$A=100$ km$^2$ s$^{-2}$ kpc$^{-1}$.  Indeed, the gas response to the
potentials is similar as expected, forming two spiral arms in both
cases.

For the specific case of a Milky Way-like galaxy, we constructed a
model that reproduces some of the observational parameters for the
background and spiral arms potential to compare them with the cosine
potential for the spiral arms (force fitted with PERLAS).  The spiral
arms pattern angular speed is $20 \kmskpc$, on a logarithmic locus
with a pitch angle of $15.5 \deg$. The mass of spiral arm for PERLAS
model is 3\% of the disk mass, which is equivalent to a cosine
amplitude of $A=650 \km^2 \sec^{-2} \kpc^{-1}$, when the
non-axisymmetric force is employed for the fitting (see
\S\ref{sec:fit}), and $A=2000\km^2\sec^{-2}\kpc^{-1}$, when the arm
strength (i.e. the $Q_T$ parameter) is considered for the fitting
instead (see \S\ref{sec:strength}). We {\bf follow the evolution of
the system} for $5\Gyr$.

\begin{figure*}

\includegraphics[width=0.4\textwidth]{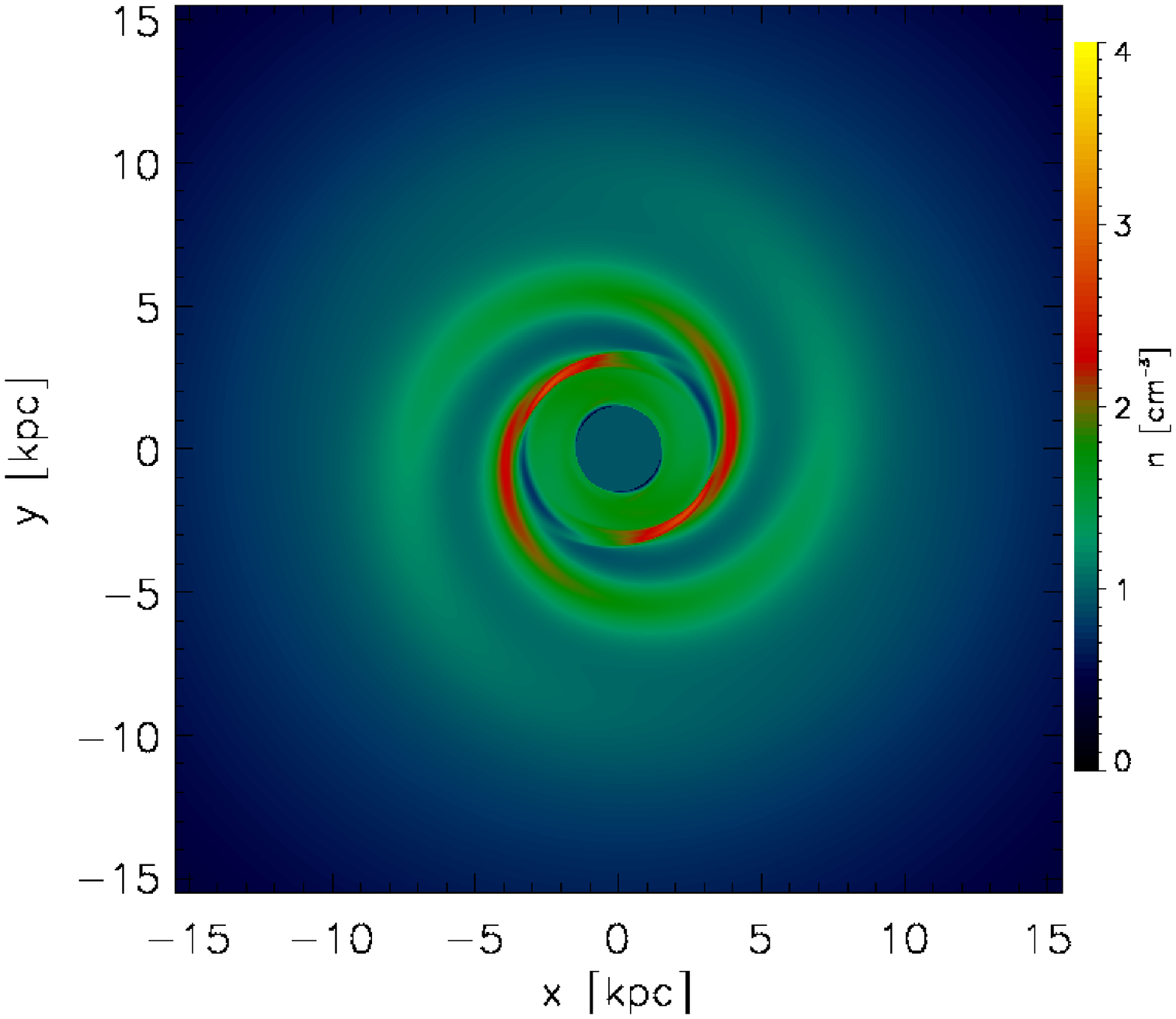}
\includegraphics[width=0.4\textwidth]{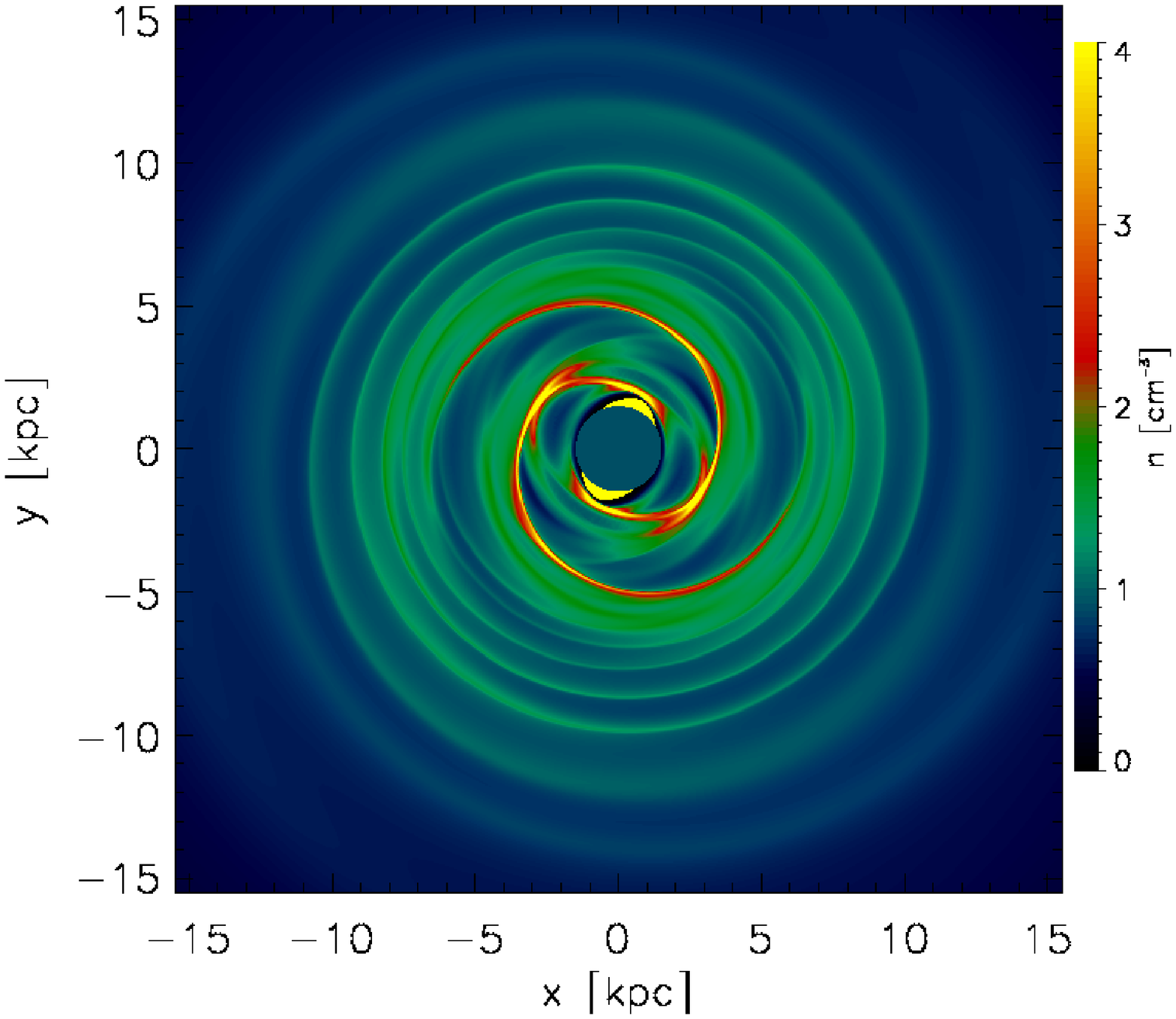}
\\
\includegraphics[width=0.4\textwidth]{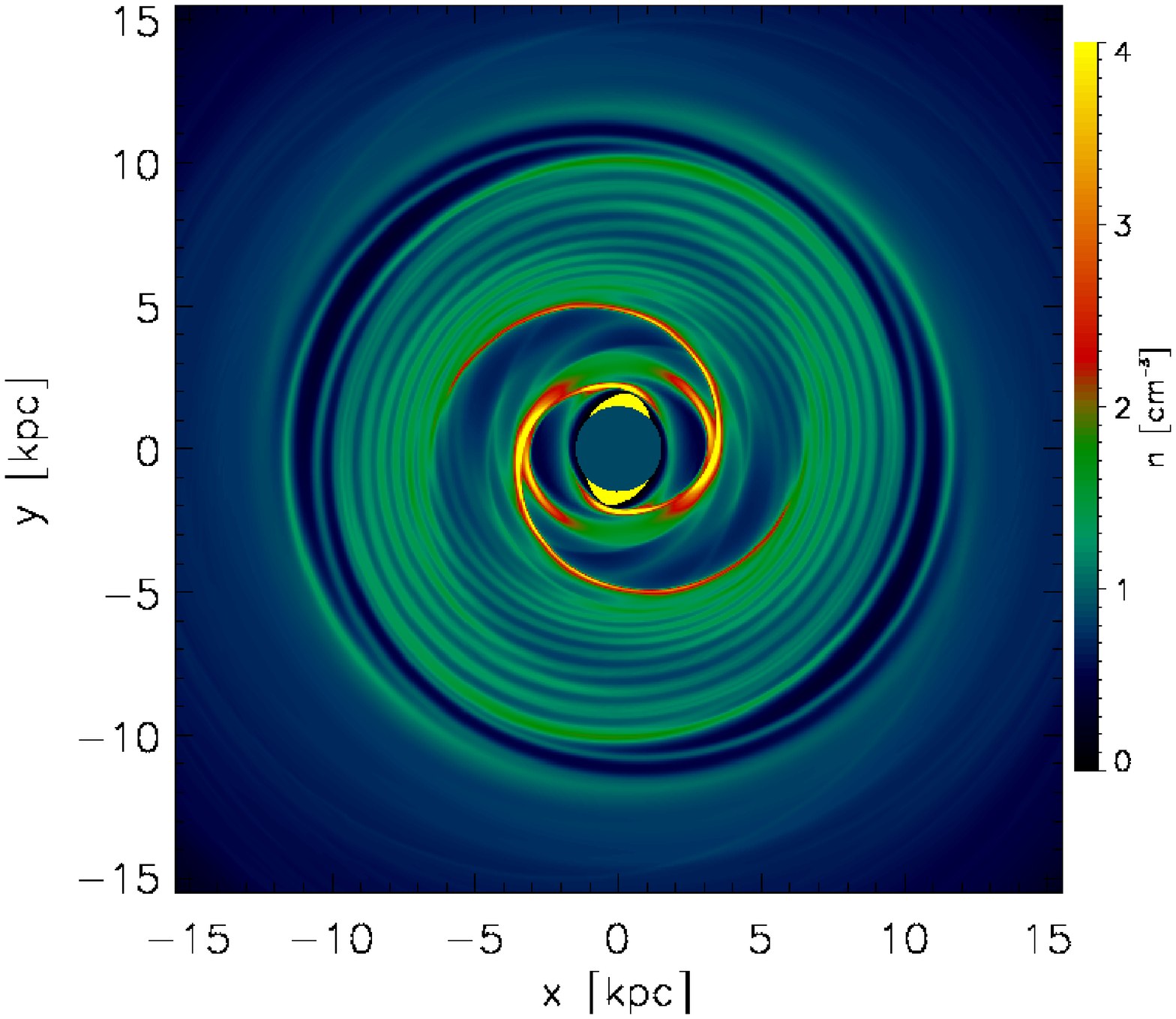}
\includegraphics[width=0.4\textwidth]{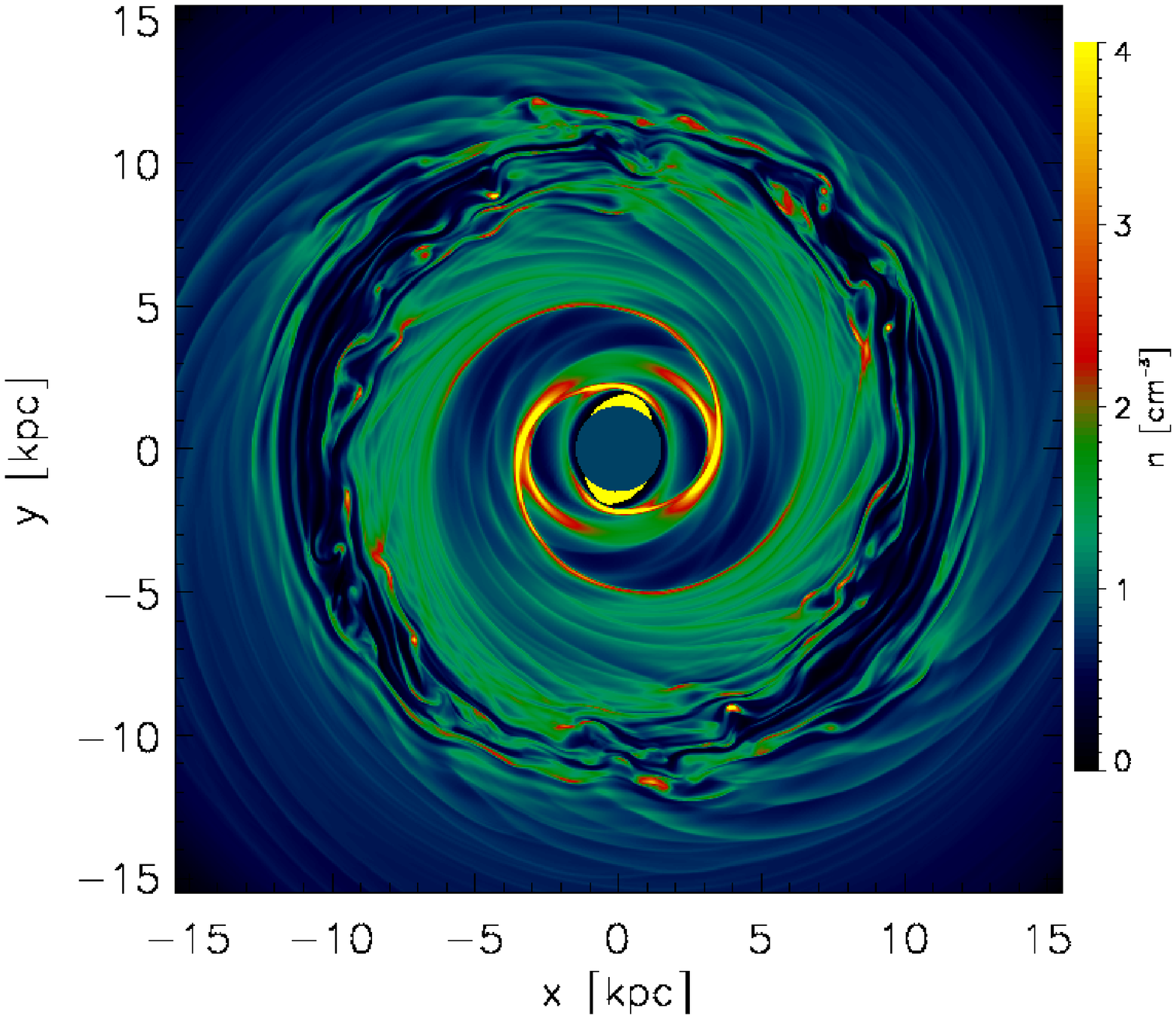}
\caption{Density distribution using a cosine potential to model spiral
  arms.  Times shown in this plot correspond to $30 \Myr$ after the
  simulation starts ({\it top left panel}), $300 \Myr$ ({\it top right
    panel}), $1.5 \Gyr$ ({\it bottom left}), and $3 \Gyr$ ({\it bottom
    right}).}
\label{fig:gas_coseno}
\end{figure*}

Figure \ref{fig:gas_coseno} shows the resulting density distribution
of the gaseous disk when is perturbed by the cosine potential with a
small amplitude ($A=650 \km^2 \sec^{-2} \kpc^{-1}$).  After the
simulation starts, the gas very rapidly settles into a spiral pattern,
forming two spiral arms at $30 \Myr$ (top left panel).  $300 \Myr$
into the simulation (top right panel), the two spiral arms are better
defined, and the gas is forming a secondary structure.  As simulation
progresses ($1.5$ and $3 \Gyr$, bottom panels), the gas continues
responding to the spiral arms and even more substructure forms, but
the large scale density response consists of two arms only.  The pitch
angle of the formed arms is $\sim 15.5\deg$, equal to the imposed
potential at the beginning of the simulation.  In the last panel, a gas
instability at the corotation radius is seen, as reported previously
\citep{Gomez_et13,Martos_et04}.

\begin{figure*}
\includegraphics[width=0.4\textwidth]{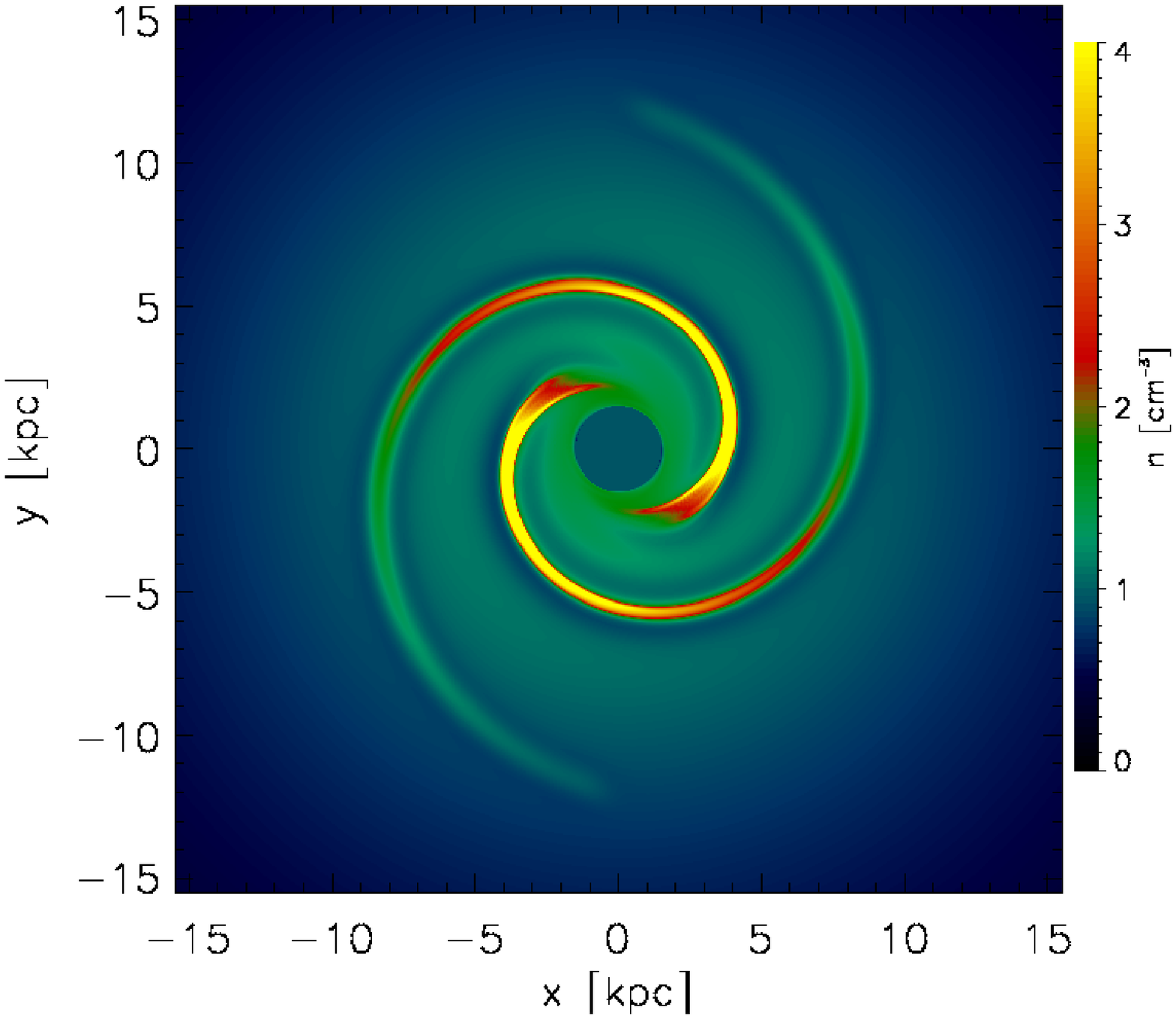}
\includegraphics[width=0.4\textwidth]{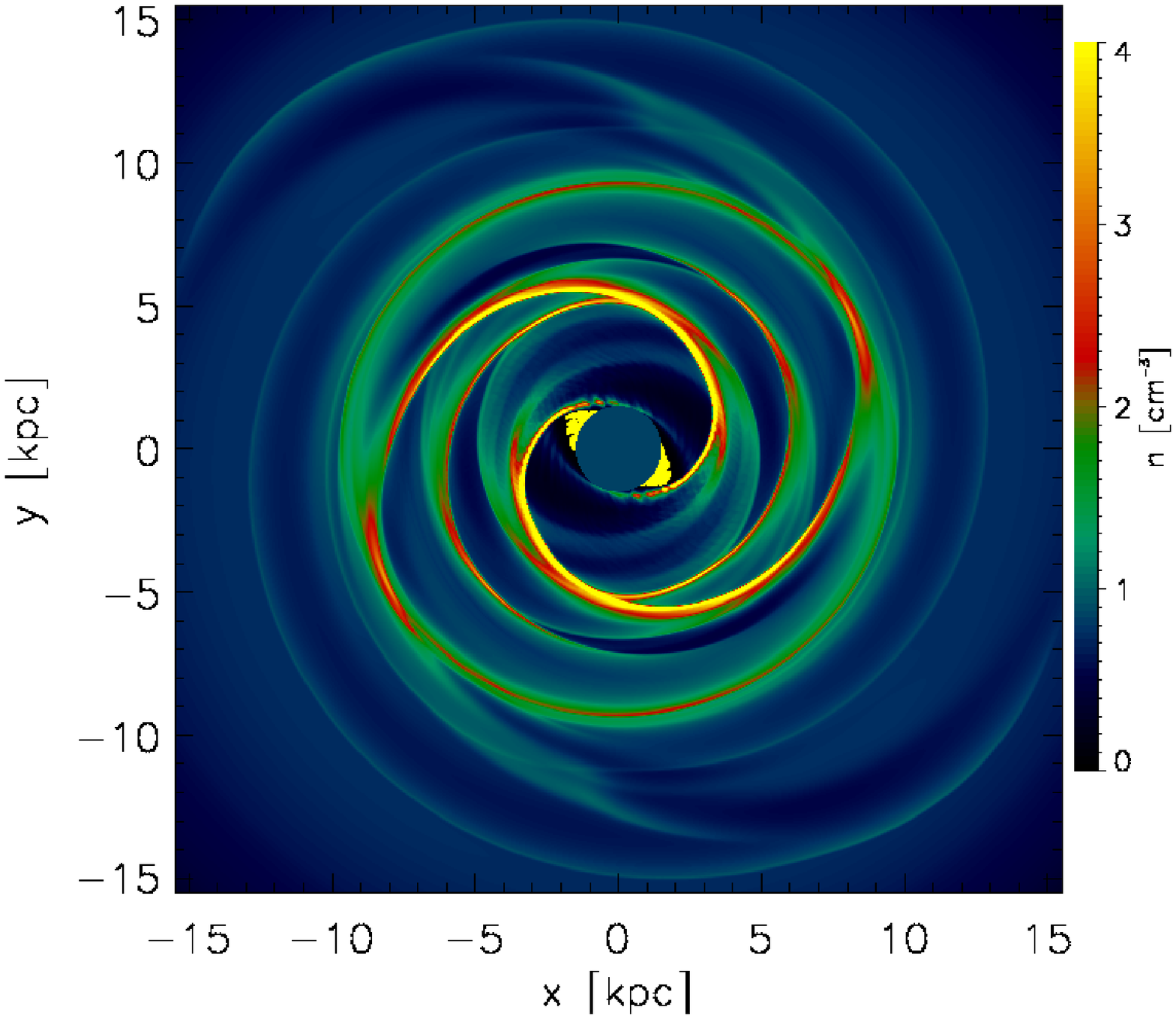}
\\
\includegraphics[width=0.4\textwidth]{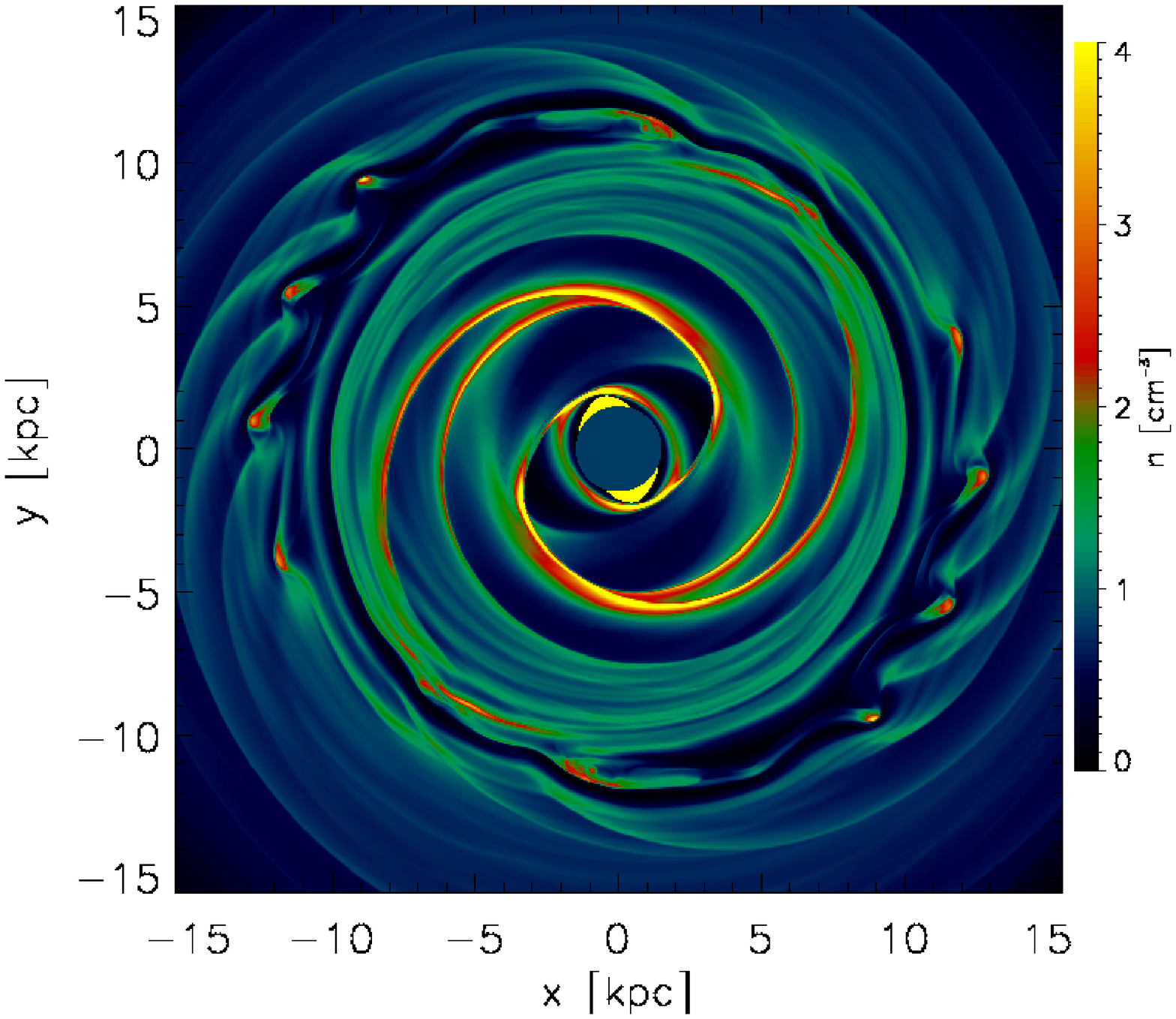}
\includegraphics[width=0.4\textwidth]{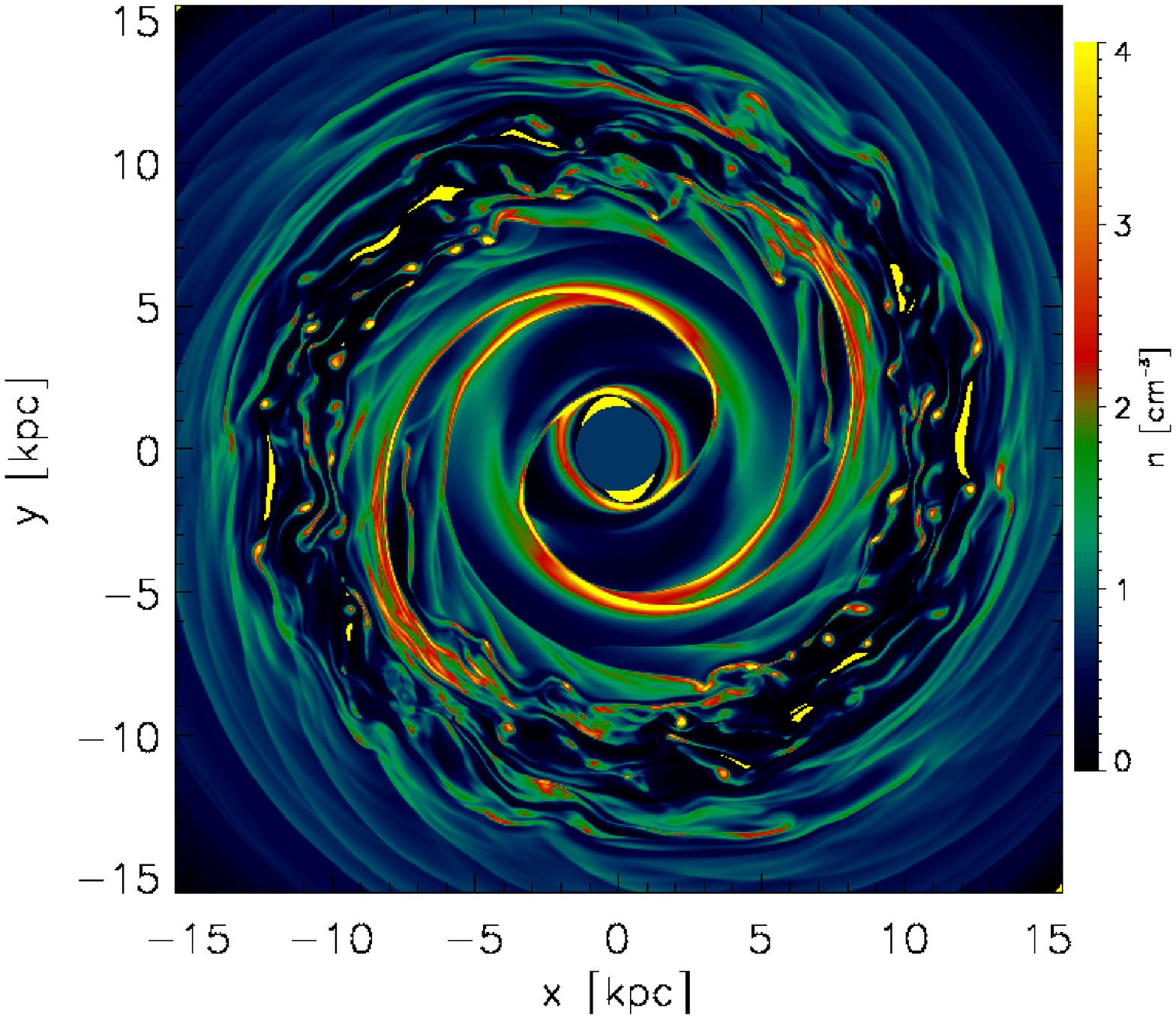}
\caption{Density distribution using the PERLAS spiral arm model.
Panels correspond to the same evolution times as in fig. \ref{fig:gas_coseno}}
\label{fig:gas_perlas}
\end{figure*}

In contrast, Figure \ref{fig:gas_perlas} shows
the density
distribution of the gaseous disk when is perturbed by the PERLAS
spiral arm potential.  The gas very rapidly settles into a spiral
pattern, as in the cosine potential case, forming two spiral arms at
30 Myr. At $300 \Myr$ and later, the gas forms four spiral arms
instead of the two arms in the cosine potential simulation. Even
though the simulation develops four spiral arms, these are associated
in two {\bf pairs, each composed of} a strong arm and a weak arm.
The strong arms
have a pitch angle of $\sim 15\deg$, and the weak arms have a pitch
angle of $\sim 7\deg$.  This doubling of the spiral arms in the gas
response has been seen in other {\bf MHD} simulations using the PERLAS model
(\citealt{Gomez_et13,Martos_et04}; see also \S\ref{sec:response}).

\begin{figure*}
\includegraphics[width=0.4\textwidth]{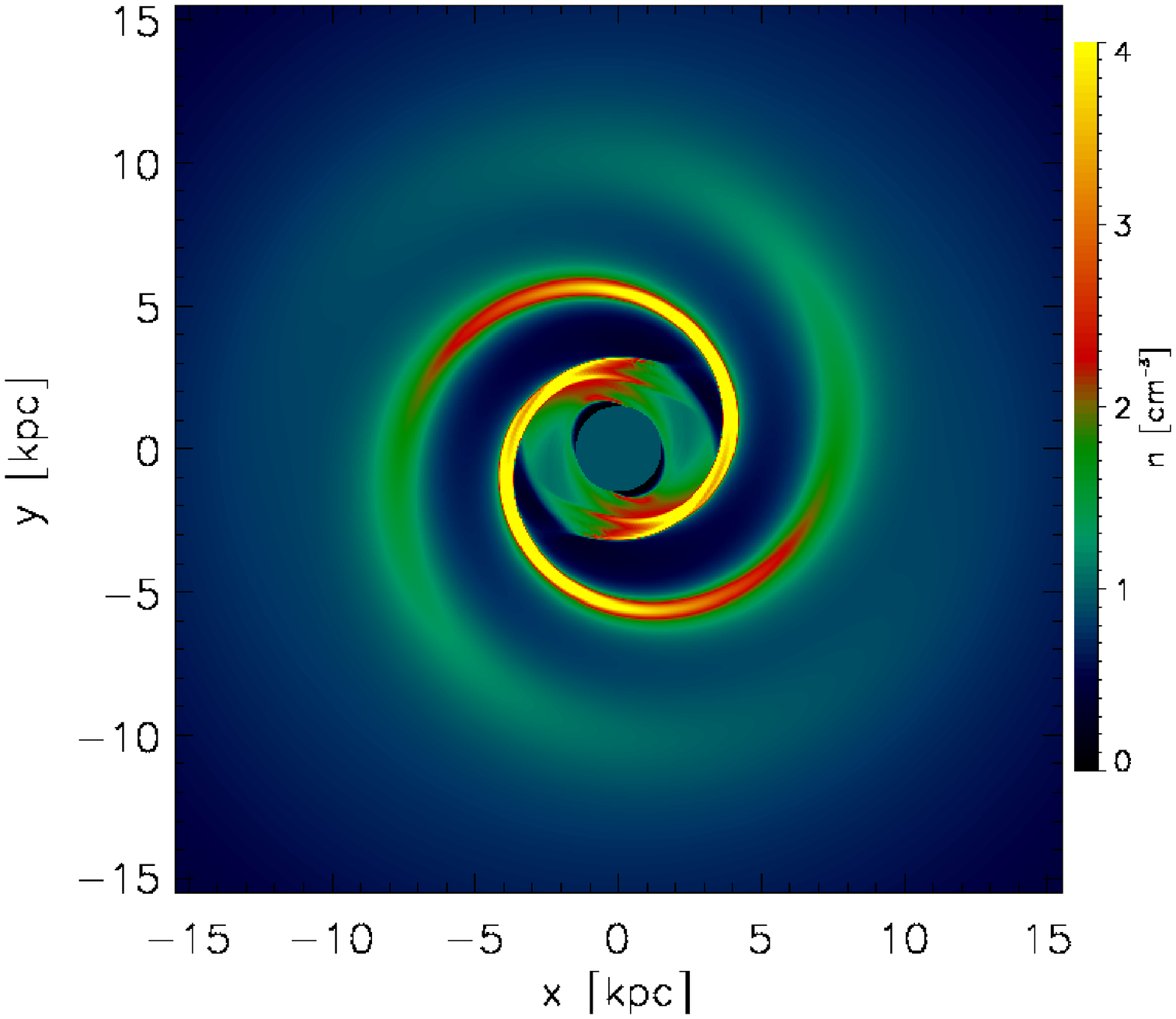}
\includegraphics[width=0.4\textwidth]{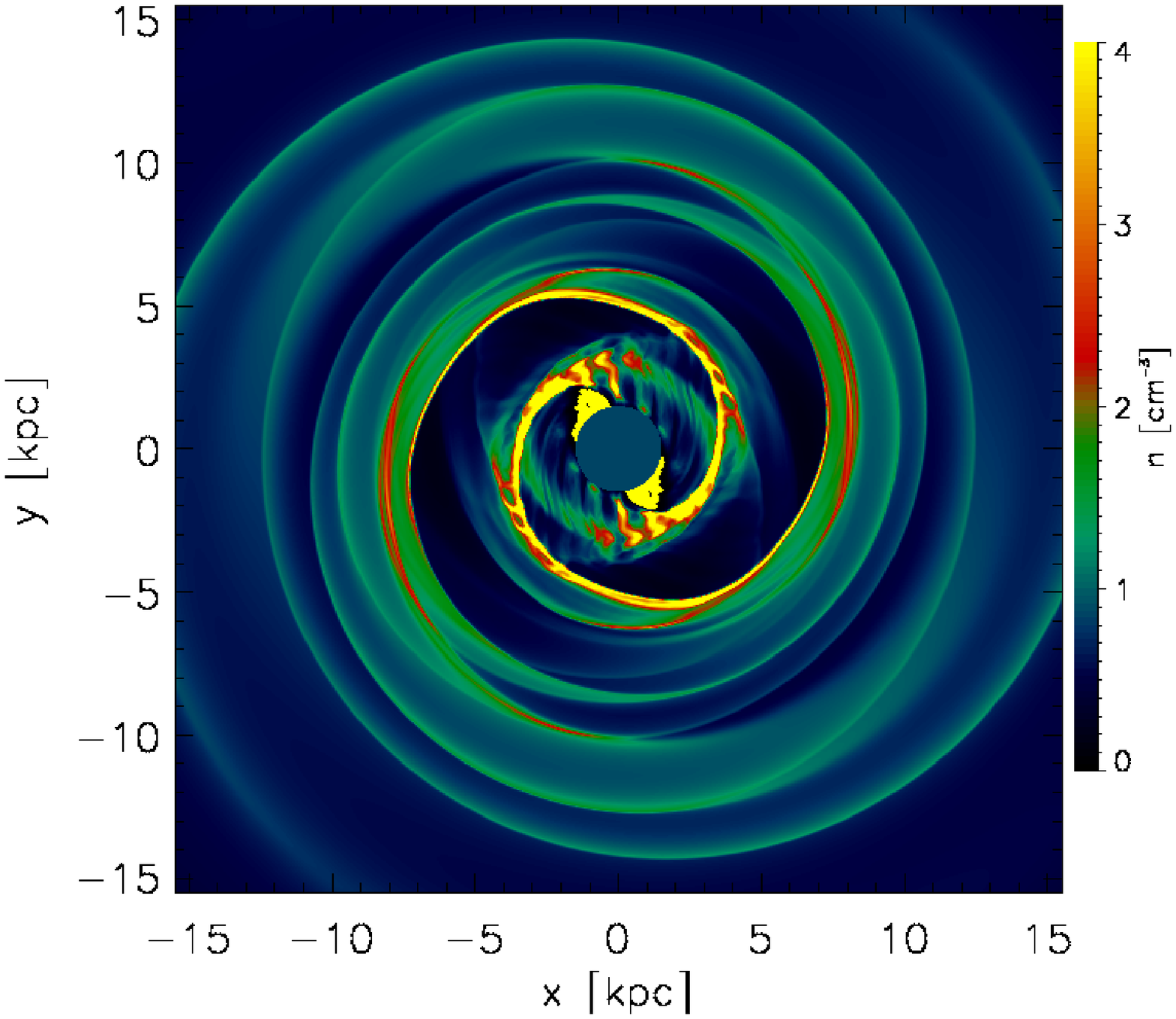}
\\
\includegraphics[width=0.4\textwidth]{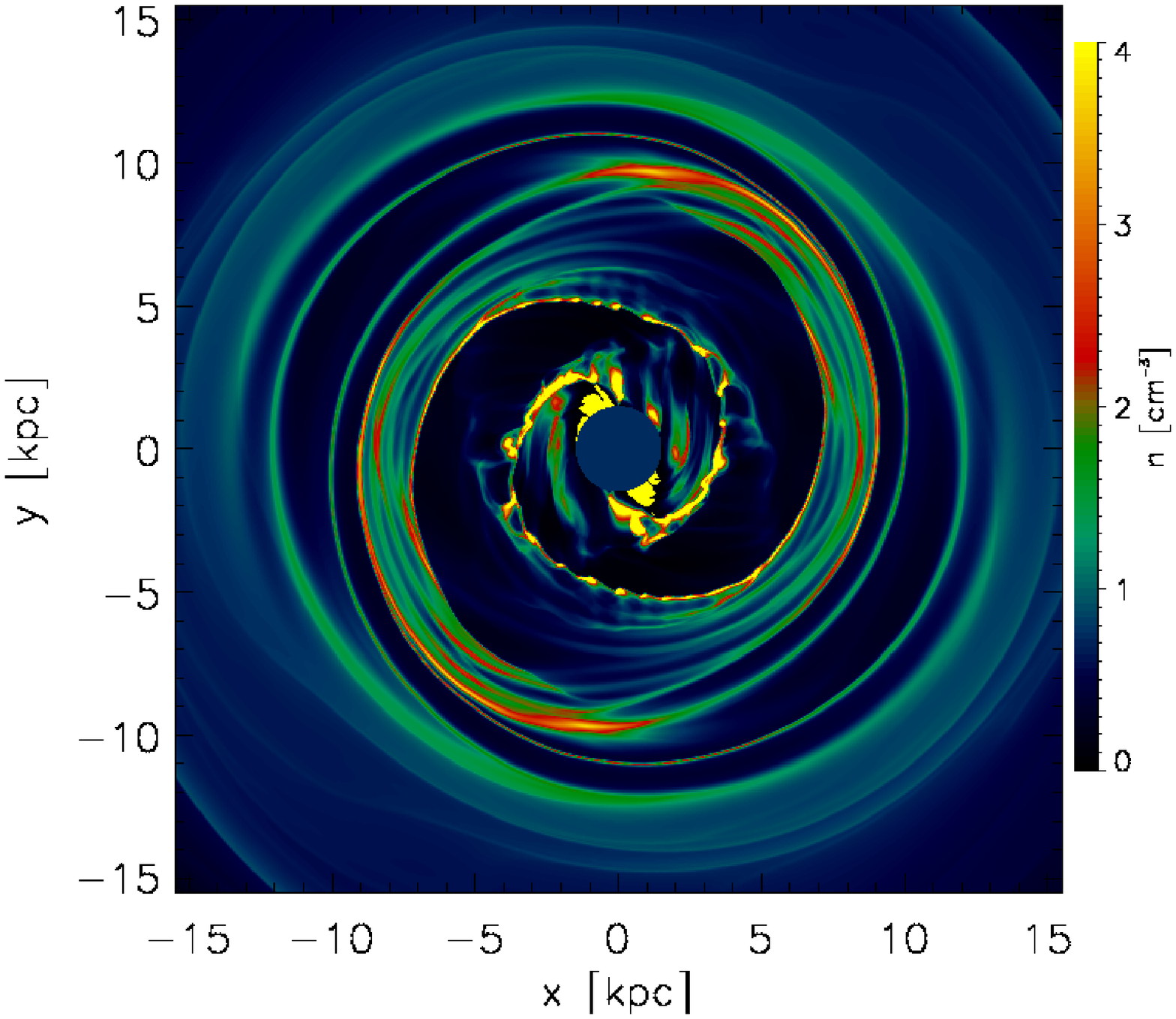}
\includegraphics[width=0.4\textwidth]{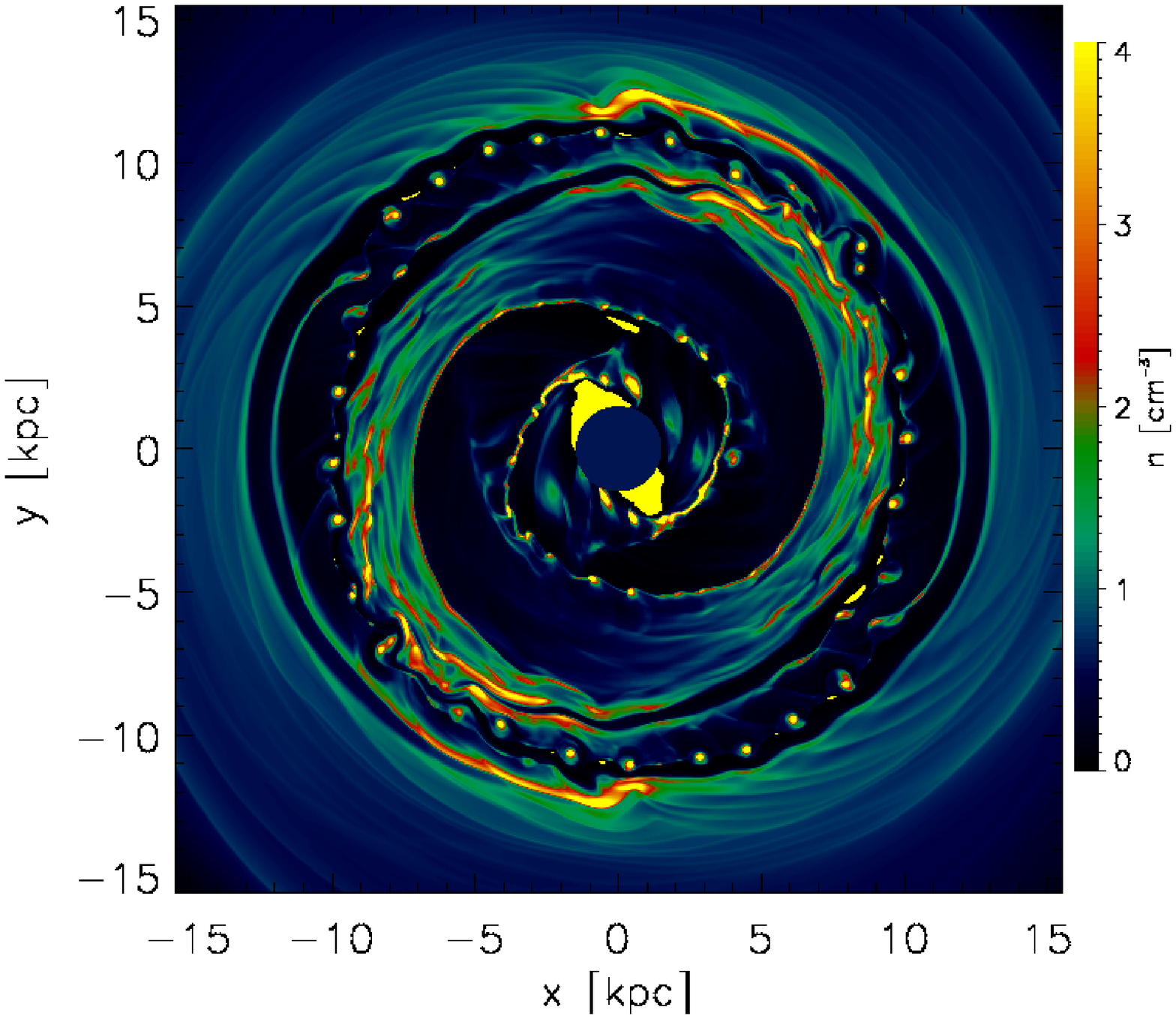}

\caption{Gas density distribution for the simulation with the cosine
  potential when its force amplitude is increased by a factor of $\sim
  3$ (up to the value provided by PERLAS model).  The times shown
  correspond to 30 Myr (top left panel), 300 Myr (top right panel),
  900 Myr (bottom left panel), and 1.5 Gyr (bottom right panel).}
\label{fig:gas_cos_aum}
\end{figure*}

In \S\ref{sec:strength}, we noted that the spiral arm strength, as
measured by eq. (\ref{eq:Q_T}), was very different for both
potentials, even if the force amplitudes were almost the
same. Therefore, we decided to perform a second experiment with a
larger amplitude $A=2000 \km^2 \sec^{-2} \kpc^{-1}$ for the cosine
potential, that would be equivalent to a factor of $\sim 3$ times the
force of PERLAS. This much larger amplitude for the cosine potential
is stronger than PERLAS at any radii, as shown in Figure
\ref{fig:QT_am20}. The purpose of this overestimated experiment was
to test if the four spiral arms formed with PERLAS were due to the
strength of spiral arms only. Figure \ref{fig:gas_cos_aum} shows the
gas density distribution resulting from perturbing the disk with the
larger force amplitude of the cosine potential. It is readily seen
that the effect of the spiral potential on the gas disk is larger and
generates more substructure, but the gas still responds forming two
spiral arms, unlike the density based potential PERLAS that forms four
arms for larger pitch angles, i.e. for stronger arms. Consequently,
this difference does not come from the force amplitude, but it seems
rather originated from the self-gravitating nature and specific
details of the potential that a local approximation for the forces
given by the cosine potential is unable to reproduce. We further
explain this in the next section.

\subsection{Branch Formation and their Relation with the Transient Nature of Spiral Arms}
\label{sec:response}

In \S\ref{sec:comparison} we showed that the intrinsic differences in
the nature of the cosine and PERLAS potentials induce a distinct
gaseous arm structure as compared to a disk perturbed by a cosine arm
potential. While employing the spiral arm cosine potential, a
bisymmetric gaseous structure seems to be an invariable outcome, for
the PERLAS potential a two or four armed structure could be
obtained. This result points to the necessity of understanding how the gas
responds to the PERLAS model when we vary the structural parameters of
the spiral arms, such as its mass and pitch angle,
considering the uncertainties in the determination of these parameters
for the Milky Way's spiral arms. Therefore, we developed a set of
simulations varying spiral arm pitch angles ($i$) and spiral arms
masses. We present here some of them ranging from $12\deg$ to
$19\deg$, and the mass of the spiral arms ($M_{sp}$), from 1.75\% to
5\% of the total disk mass ($M_D$).

\begin{figure*}
\includegraphics[width=0.27\textwidth]{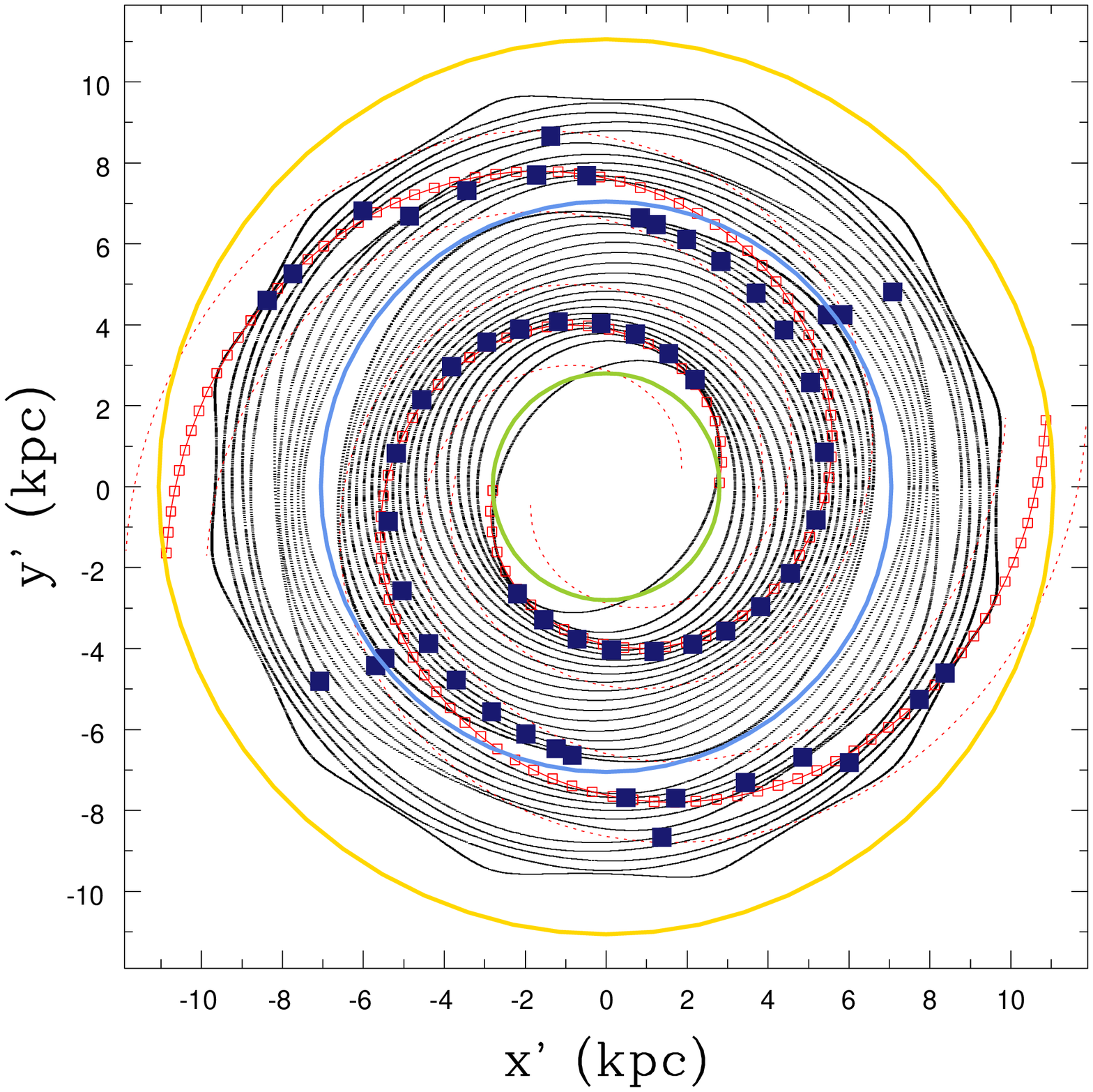}
\includegraphics[width=0.31\textwidth]{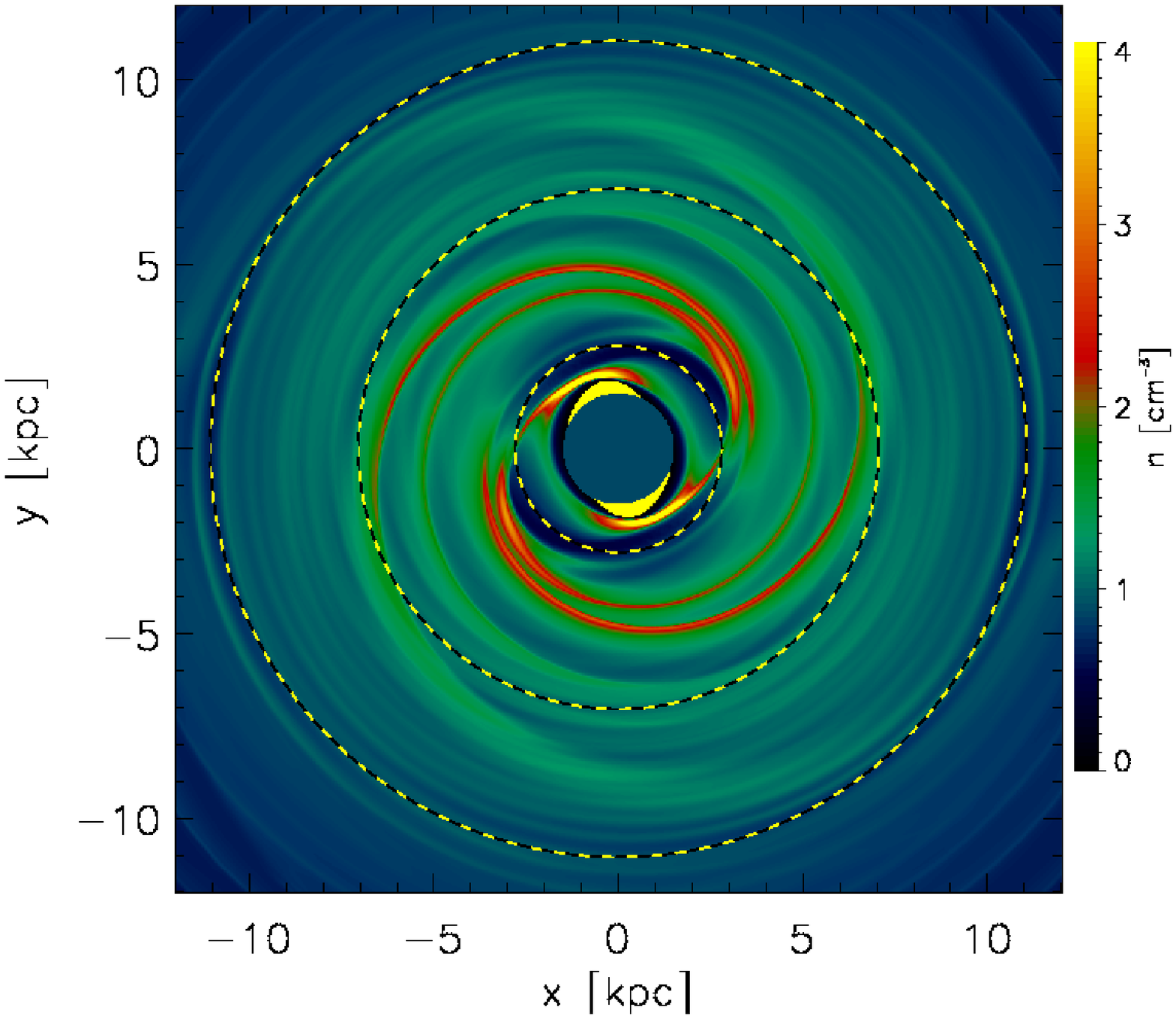}
\\
\includegraphics[width=0.27\textwidth]{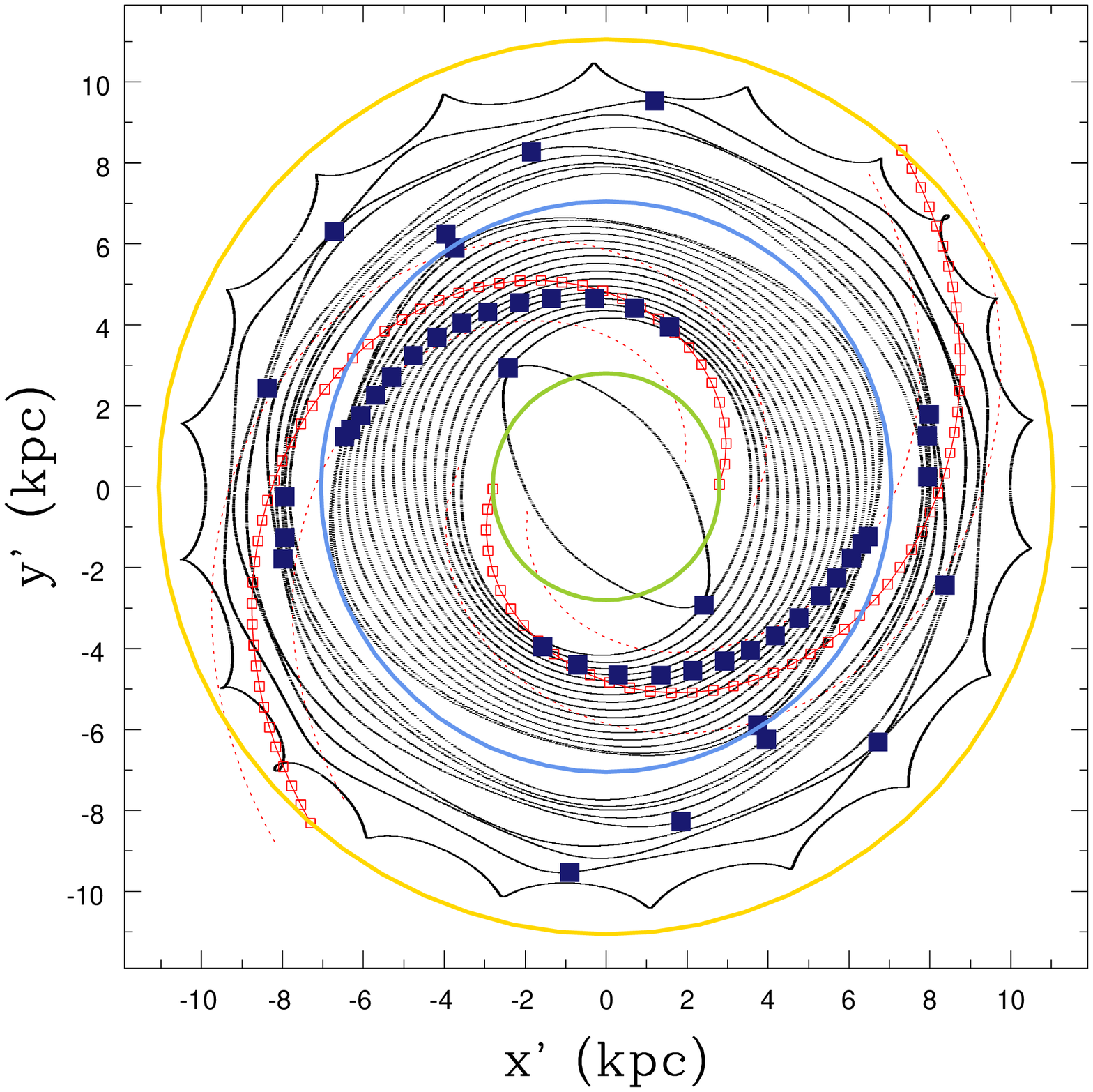}
\includegraphics[width=0.31\textwidth]{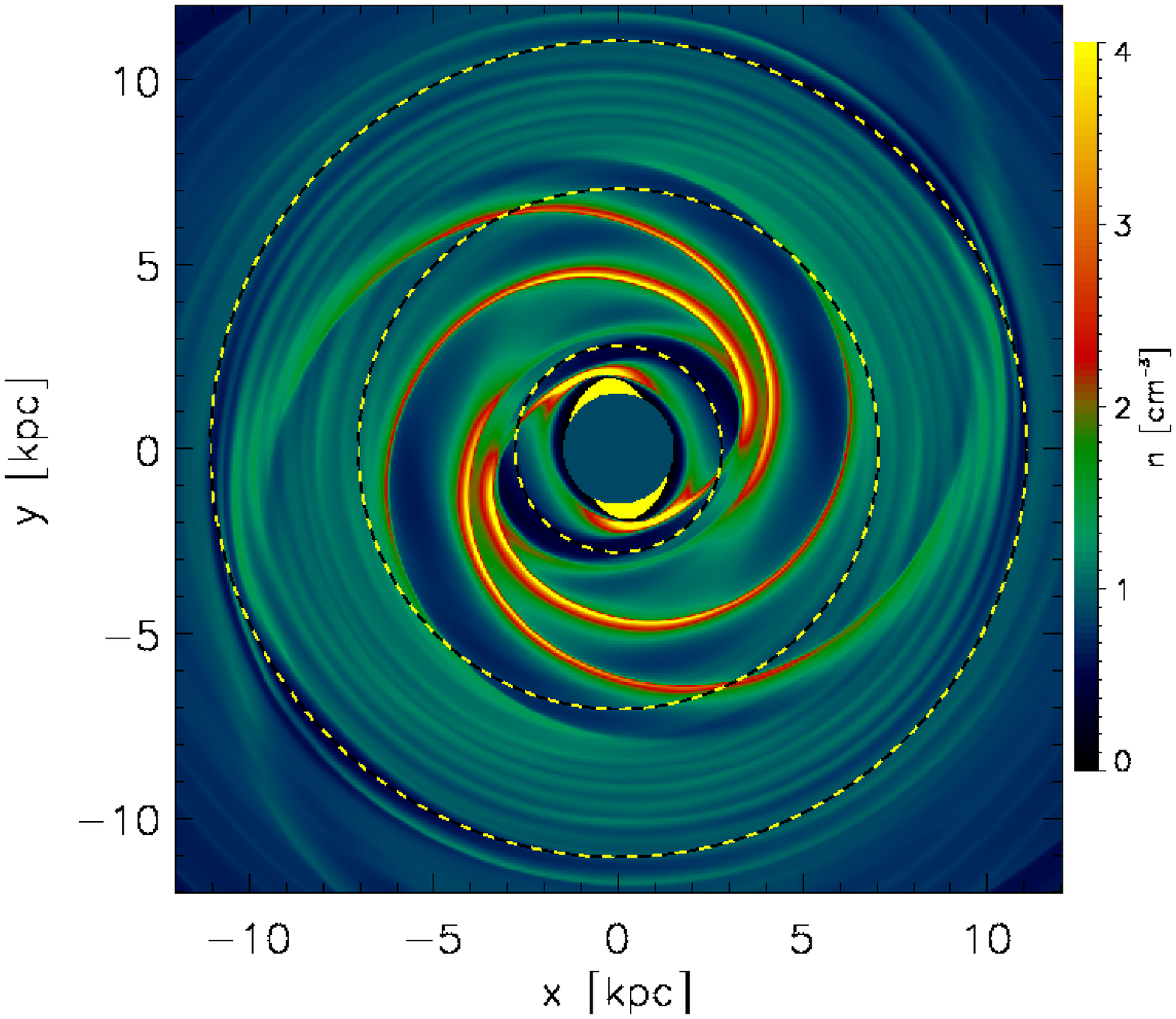}
\\
\includegraphics[width=0.27\textwidth]{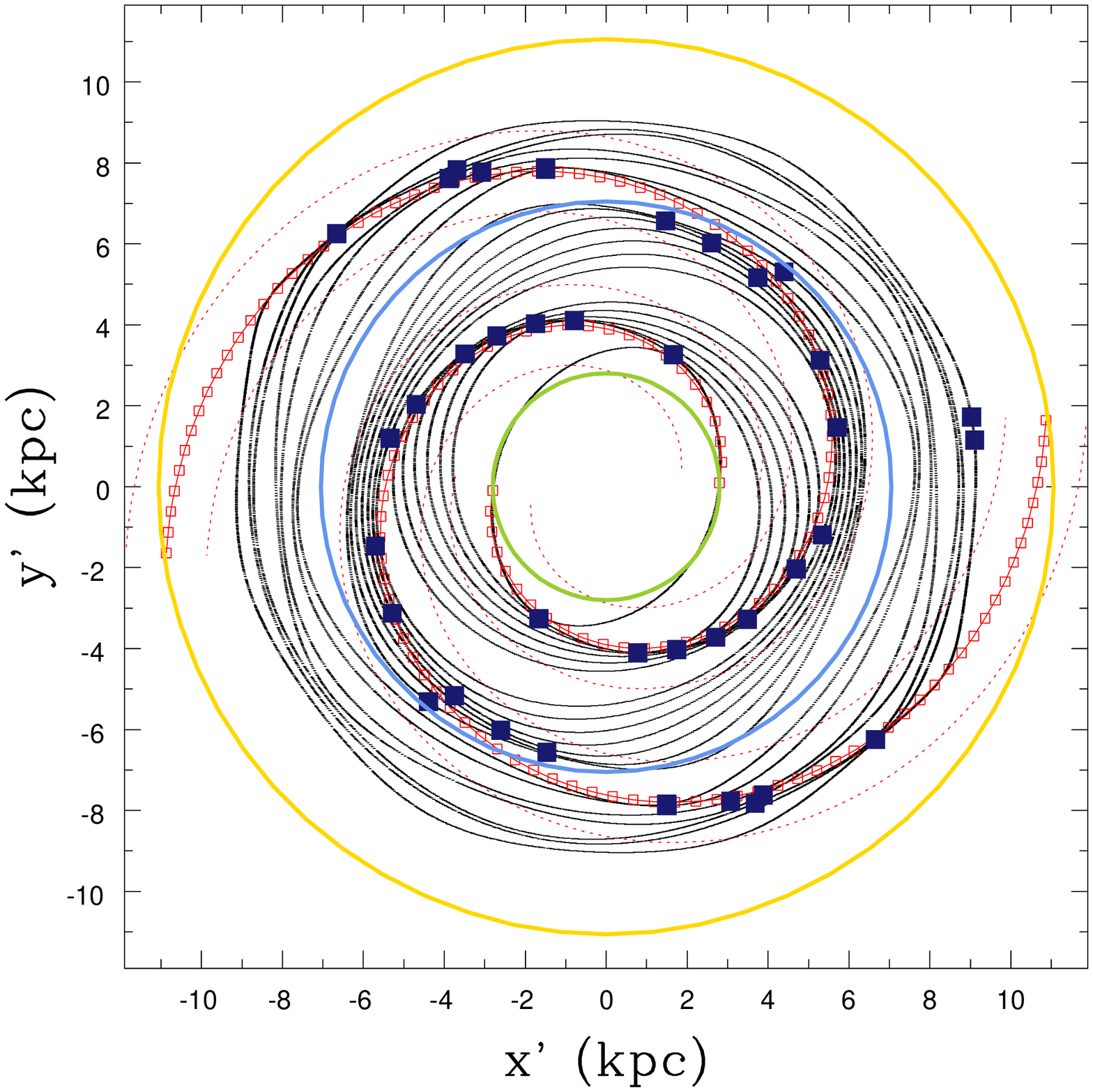}
\includegraphics[width=0.31\textwidth]{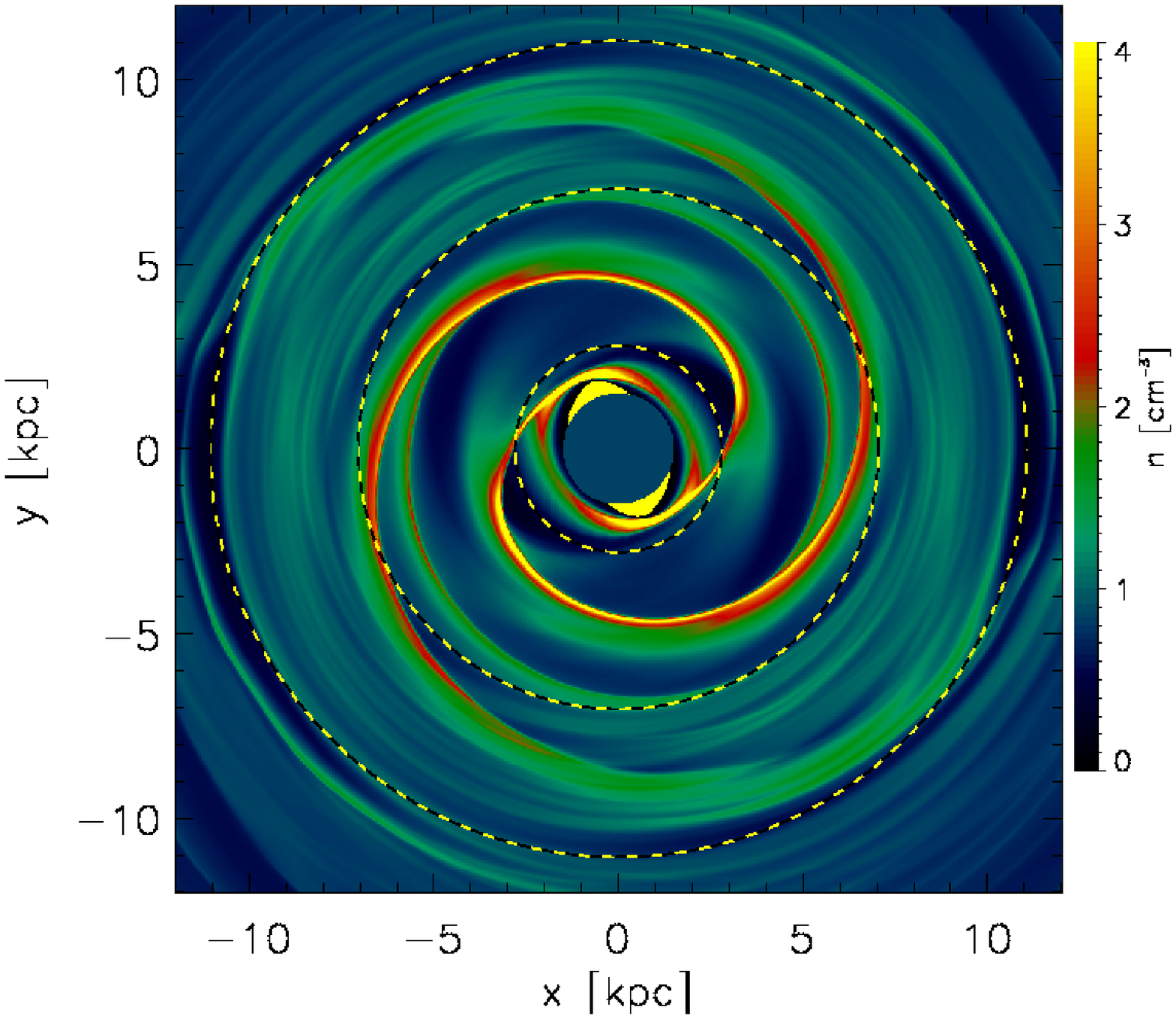}
\\
\includegraphics[width=0.27\textwidth]{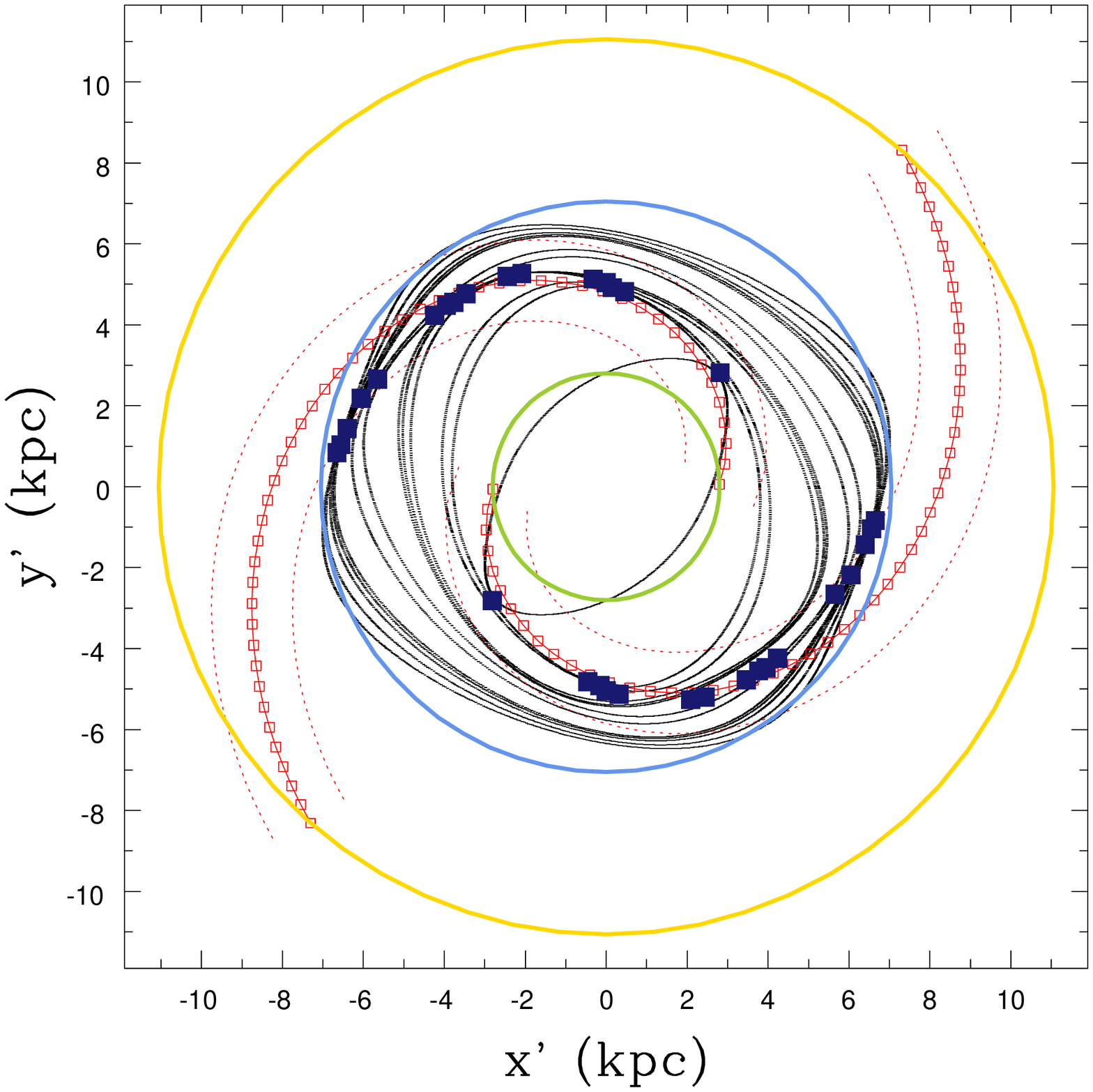}
\includegraphics[width=0.31\textwidth]{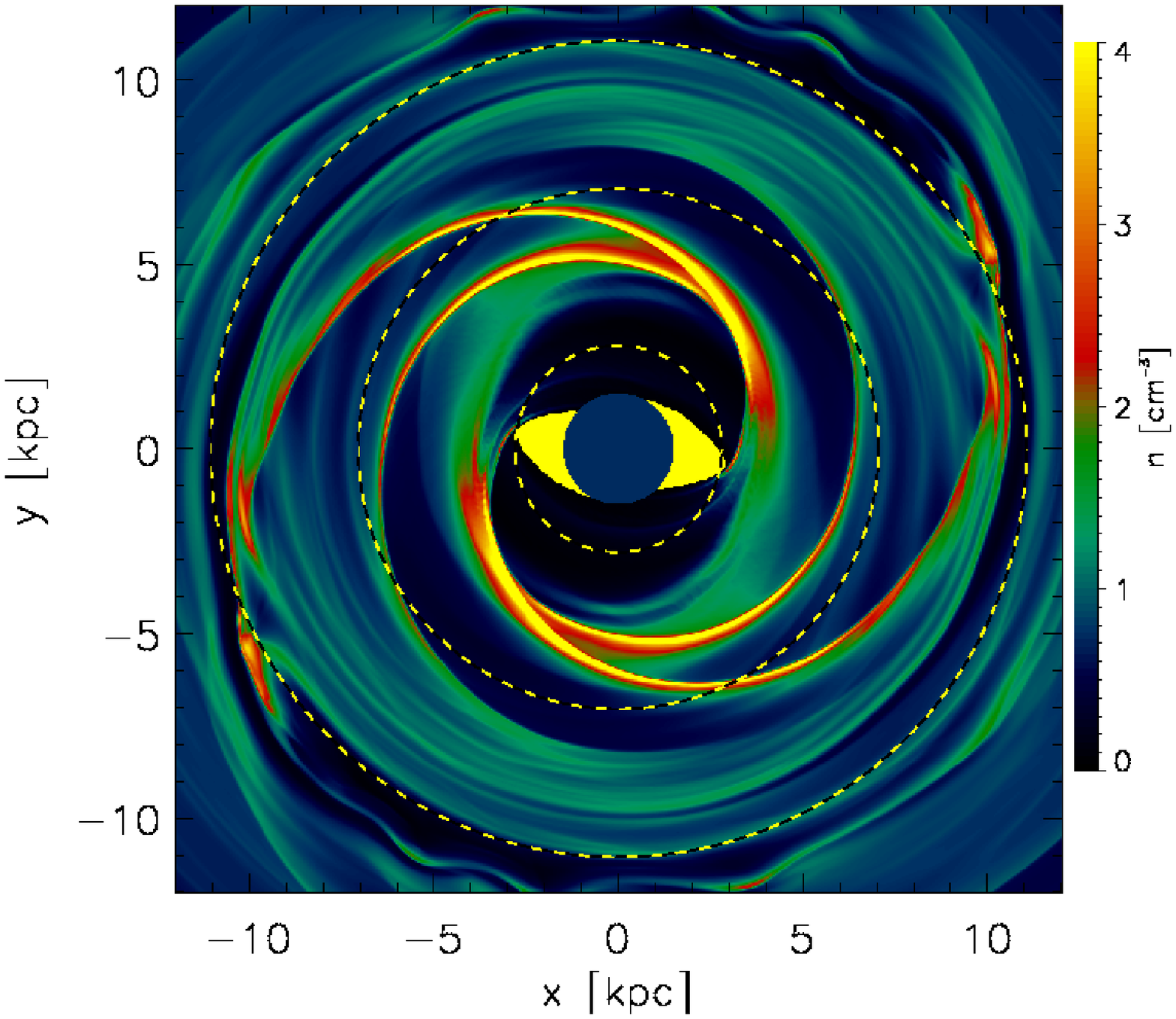}
\caption{Stellar and gaseous response to different parameters of the
  spiral arm model (PERLAS). Panels on the left column show the stable
  periodic stellar orbits ({\it black solid lines}), the response
  density maxima ({\it filled squares}), and the imposed spiral arm
  locus ({\it open squares}), flanked by two dotted lines that
  represent the spiral arm width. Colored circles represent the
  position of the inner Lindblad ({\it green}), 4/1 ({\it blue}) and
  corotation ({\it yellow}) resonances.  Panels on the right column
  show the gas density distribution after $1.2\Gyr$, together with the
  inner Lindblad, 4/1 and corotation resonances ({\it dashed
    circles}).  The mass of spiral arms and pitch angle are: $M_{sp}=
  0.0175M_D$ with $i=12\deg$ ({\it first row}) and $i=19\deg$ ({\it
    second row}), $M_{sp}= 0.05M_D$ with $i=12\deg$ ({\it third row})
  and $i=19\deg$ ({\it fourth row}).}
\label{fig:periodic_gas}
\end{figure*}

Figure \ref{fig:periodic_gas} shows a mosaic of simulations. The
panels in the left column are the stable periodic stellar orbits and
the panels in the right column are the gas density distribution after
$1.2\Gyr$. {\bf The maximum density response (where the orbits crowd 
producing a density enhancement) quantifies the orbital support to a 
given spiral perturbation through periodic orbits. We computed the stellar 
periodic orbits in order to explore the orbital support to the imposed PERLAS spiral arm potential (open
squares in the left column-panels of Figure \ref{fig:periodic_gas}).  
To estimate the density response, we employ the method of \citet{CG86}.
This method assumes that stars in circular orbits in an axisymmetric potential, 
with the same sense of rotation of the spiral arms, are trapped around the 
corresponding periodic orbit in the presence of the spiral arms. For this purpose, 
we calculated a set of central periodic orbits (between 50 and 60) and found the 
density response along them using the conservation of the mass flux between any 
two successive periodic orbits. With this information we seek the position of the 
maximum density response along each periodic orbit (filled squares in the left 
column-panels of fig. \ref{fig:periodic_gas}).  These positions are compared
with the imposed spiral arms (PERLAS model). The method implicitly considers a small 
dispersion (with respect to the central periodic orbit) since it studies a 
region where the flux is conserved. On the other hand, this dispersion is based on 
parameters for the galaxies where dynamics is quite ordered, orbits follow their periodic orbit closely, 
in such a way that we consider this study a good approximation. This method to estimate the 
density response has been widely used in literature \citep{ConG88, AL97, Yano03,PMME03,VSK06,TEV08, 
Perez-Villegas_et12, Perez-Villegas_et13,JunLep13}. We refer the
reader to the work of \citet{CG86} for more details.}
{\bf With this in mind, a model in Figure \ref{fig:periodic_gas}} where
the open and filled squares coincide would represent a stable,
approximately orbitally self-consistent system, while a lack of
coincidence would mean that the spiral is unlikely {\bf to be} long-lasting
\citep{Perez-Villegas_et12,Perez-Villegas_et13}.
 
For $M_{sp}=0.0175M_D$ and $i=12\deg$ (upper left panel of Figure
\ref{fig:periodic_gas}), the stellar density response follows
approximately the imposed spiral potential prior to the 4/1
resonance. After that, the stellar response forms a slightly smaller
pitch angle than the imposed spiral. In the MHD simulation, the
gaseous disk responds to the two-arm potential with the now familiar
four spiral armed structure, where the pitch angle of the stronger
pair of arms corresponds to that of the imposed pitch angle potential,
while the other pair of gaseous arms has a systematically smaller
pitch angle, corresponding closely to the regions of periodic orbits
crowding. Now, with the same spiral arm mass but a pitch angle of
$i=19\deg$ (second row of fig. \ref{fig:periodic_gas}), the stellar
density response (i.e. periodic orbits) forms again a smaller pitch
angle compared with the imposed, while in the gas, the four spiral
arms seem stronger and the difference in the pitch angle between the
gaseous and imposed arms is larger than the previous case. For
$M_{sp}=0.05M_D$ and $i=12\deg$ (third row), the stellar density
response closely follows the imposed spiral arm potential prior to the
4/1 resonance. After that, the stellar response forms a slightly
smaller pitch angle than the imposed spiral, while the gas responds
with four spiral arms, but the second pair is very weak but with a
significantly smaller pitch angle than the imposed spiral arm
potential. Finally, with the same spiral arm mass (5\% of the disk
mass) but a pitch angle of $i=19\deg$ (fourth row),
we found
no periodic orbits beyond 4/1 resonance and the stellar density
response forms a pitch angle smaller than $19\deg$.
The gaseous disk
responds with well defined four spiral arms that extend up to the
corotation radius. Notice that in this last simulation, there is not
much that can be said about the (stellar or gaseous) orbital support
since periodic orbits tend to disappear due to the strong forcing of
the imposed spiral arms, meaning that spiral arms would rather be
transient by construction in this case and the MHD gaseous disk
behavior is difficult to predict from periodic orbit
computations. However, such as the stellar arms constructed in this
case, the gaseous spiral would be transient in likely even shorter
timescales than in the case where periodic orbits exist but settle
down systematically in smaller pitch angles than the original imposed
spiral arms in the region where periodic orbits do exist. Notice
that, in
general, the arms that should eventually disappear in this scenario,
are the stronger stellar imposed spiral arms (see second row of fig.
\ref{fig:periodic_gas}; \citealt{Perez-Villegas_et13}).

Therefore, the stellar response density maxima represent the regions
of the arms where stars would crowd for long timescales, this is,
where the existence of stable, long-lasting spiral arms would be more
likely. On the other hand, if the stellar density response forms a
spiral arm with a different pitch than the imposed angle,
then the imposed spiral arms triggered on the disk (by a bar, an
interaction, etc.), would rather be structures of transient nature
since those are not orbitally supported \citep{Perez-Villegas_et13}.
Likewise, for the case of the spiral arms in the Milky Way, the values
frequently seen in literature for the pitch angle, range from $\sim
11\deg$ to $19\deg$. These values and the knowledge of the galactic type
could provide some information about their nature, i.e. whether
they are long-lasting or transient structures.
{\bf Following the pitch
angle restrictions found by \citet{Perez-Villegas_et13},
the larger pitch angle values reported in literature for the Milky
Way galaxy would imply that its spiral arms are
a transient feature.}
The formation of four spiral arms
in the gas response is, in this scenario, another piece of evidence of
a transient nature of the spiral arms in the Milky Way galaxy, as we
claim it is the secondary pair of arms in Figure
\ref{fig:periodic_gas}, that coincides {\bf more} closely with the stellar
density response {\bf from} periodic orbits, as expected from
\citet{Gomez_et13}.
Summarizing, the
first (imposed) pair of massive spiral arms formed in the disk, with a
larger pitch angle, triggers a second pair of arms
(traced approximately by the
periodic orbits), with smaller pitch angles
\citep{Perez-Villegas_et13}. In this outline, the gas responds
forming a second pair of arms aligned with the locus of the orbital
crowding.
This lighter structures would likely be
preferentially formed by young stars and gas than by evolved stars
because of their transient nature, i.e. similar to what we call
``branches''. Finally, in this framework, the presence of clear and
strong branches in spiral galaxies, with smaller pitch angles than the
corresponding couple of massive spiral arms on a galaxy, would be a
signature of the transient nature of the spiral arms in a given
galaxy. On the other hand, spiral galaxies without evidence of
branches could indicate the presence of a longer lasting spiral arm
structure.

\section{Discussion and Conclusions}\label{sec:conclusions}

With the use of a detailed three dimensional, density-distribution
based potential for the spiral arms, combined with MHD simulations on
a Milky Way-like galactic disk, we have studied the stellar orbital
and gaseous response to the galactic potential. As a first experiment,
we constructed a simple cosine potential (as the ones
commonly employed in literature) that reproduced approximately what
the density based potential PERLAS exerts on the stellar and gaseous
dynamics.

The first set of simulations compare the gas response when the disk is
perturbed by both spiral arm potential models. We also varied the
structural parameters of the spiral arms in the PERLAS model, such as
the pitch angle ($12\deg$ to $19\deg$) and the mass of spiral arms
(1.75\% to 5\% of the stellar disk mass) in order to understand how
these parameters affect the gaseous disk dynamics. Additionally, we
constructed stellar periodic orbits and calculate the stellar response
density maxima. 

With these exercises we found that only in the case of the PERLAS
model
the gas and stellar density response (based on the existence of
periodic orbits) is a
consistent four-armed spiral structure: a couple of strong gaseous arms
located at the position of the imposed stellar arms, and a second pair of
weaker gaseous arms located at the position of the stellar orbit crowding.
{\bf Since the potential and the stellar density response do not coincide,
the spiral arms are}
prone to destruction.
{\bf So, the presence of the second (gaseous) pair is interpreted as
a sign of this transiency.}

Our main conclusions can be summarized as follows:

\begin{itemize}
  \item
    {\bf We performed a}
    study of the gas response to {\bf two} galactic
    potentials: the density distribution based model PERLAS and the widely
    employed in literature cosine potential.
    We verified that the gas response
    (a two armed structure) is the same for both
    models close to the linear regime only,
    i.e. {\bf for low-mass spiral arms and
    pitch angles smaller than $\sim 10\deg$.}
    In the
    general case however (i.e. large pitch angle or arm mass), even when
    the cosine potential has been fitted as close as possible
    to the PERLAS model, they produce quite distinct outcomes on
    the gas response. In the case of the cosine potential, the gas
    responds invariably forming two spiral arms while, with the PERLAS
    model, the gas responds with four gaseous spiral arms (in the
    general strong arm case).

  \item
    We increased the strength of the spiral arms represented by the cosine
    potential up to a point the arms were equivalent and even beyond
    the mass of a strong bar
    as an experiment to try to reproduce the gas response provided by the
    PERLAS model.
    However, the answer was always a bisymmetric structure.
    We conclude that the spiral arm strength
    is not responsible for the
    four-arm gas response, but rather it is the product of the
    forcing generated by the whole density distribution
    better represented by the PERLAS model that, in turn, forces the periodic
    orbit response to shift its crowding regions inside the imposed
    locus of the massive spiral arms.

  \item
    Using the PERLAS model, we changed the structural parameters of the
    spiral arms according to the observational and theoretical
    uncertainties in the determination of the Milky Way's spiral arms in
    literature. Within these values, for the general case, the stellar
    response density maxima systematically forms spiral arms with a
    smaller pitch angle than the imposed spiral, meaning that the spiral
    arms might be a transient feature in the Milky Way Galaxy
    \citep{Perez-Villegas_et13}. The presence of a second pair of lighter
    spiral arms (``branches''), with smaller pitch angles induced by the
    first more massive stellar spiral arms, might be evidence
    of the lack of support to the stronger arms on a galactic
    disk, and therefore evidence of the transient nature of spiral arms in
    a given galaxy. Applying this scenario to the Milky Way, for
    the stronger spiral arm values reported in literature (i.e. pitch angles
    larger than $\sim 10\deg$, and masses larger than $\sim 2\%$ of the
    disk), the spiral arms in the Milky Way would be of transient nature.

  \item
    Although in this work we applied the models to Milky Way-like
    galactic discs, it is worth noticing that the results are general.
    This means that the
    presence of branches with smaller pitch angles than the main
    arms might be the ``smoking gun'' that proves transiency of spiral
    arms in any given galaxy.

\end{itemize}

\section*{Acknowledgments}

We thank Edmundo Moreno for enlightening discussions that helped to
improve this work. APV acknowledges the support of the postdoctoral
Fellowship of DGAPA-UNAM, M\'exico. This work has received financial
support from DGAPA PAPIIT grants IN111313 and IN114114.

\label{lastpage}

\end{document}